\documentclass[aps,prd,reprint,superscriptaddress,twocolumn]{revtex4-2}

\usepackage{comment}
\usepackage[utf8]{inputenc} 
\usepackage{graphicx,color,overpic,mathtools}
\usepackage{amsthm,amsmath,amssymb,hyperref,mathrsfs}
\usepackage{flexisym}
\usepackage{braket,bm,bbm,setspace}
\PassOptionsToPackage{normalem}{ulem}
\usepackage{ulem} 
\usepackage{physics}
\usepackage{float}
\usepackage[makeroom]{cancel}
\usepackage[english]{babel}
\usepackage{graphicx}
\graphicspath{ {images/} }
\usepackage{breqn}
\usepackage{bm}
\addto\captionsspanish{}
\hypersetup{
    colorlinks=false,
    pdfborder={0 0 0},
}
\usepackage[usenames,dvipsnames]{xcolor}

\makeatletter
\let\cat@comma@active\@empty
\makeatother

\begin{document}

\title{Asymptotically flat vacuum solutions in order-reduced semiclassical gravity}

\author{Julio Arrechea}
\author{Carlos Barcel\'o} 
\affiliation{Instituto de Astrof\'isica de Andaluc\'ia (IAA-CSIC),
Glorieta de la Astronom\'ia, 18008 Granada, Spain}
\author{Ra\'ul Carballo-Rubio}
\affiliation{CP3-Origins, University of Southern Denmark, Campusvej 55, DK-5230 Odense M, Denmark}
\affiliation{Florida Space Institute, University of Central Florida, 12354 Research Parkway, Partnership 1, 32826 Orlando, FL, USA}
\author{Luis J. Garay} 
\affiliation{Departamento de F\'{\i}sica Te\'orica and IPARCOS, Universidad Complutense de Madrid, 28040 Madrid, Spain}

\begin{abstract}
We investigate the effects of quantum backreaction on the Schwarzschild geometry in the semiclassical approximation. The renormalized stress-energy tensor (RSET) of a scalar field is modelled via an order reduction of the analytical approximation derived by Anderson, Hiscock and Samuel (AHS). As the resulting AHS semiclassical Einstein equations are of fourth-derivative order in the metric, we follow a reduction of order prescription to shrink the space of solutions. Motivated by this prescription, we develop a method that allows to obtain a novel analytic approximation for the RSET that exhibits all the desired properties for a well-posed RSET: conservation, regularity, and correct estimation of vacuum-state contributions.
We derive a set of semiclassical equations sourced by the order-reduced AHS-RSET in the Boulware state. We classify the self-consistent solutions to this set of field equations, discuss their main features and address how well they resemble the solutions of the higher-order semiclassical theory. Finally, we establish a comparison with previous results in the literature obtained through the Polyakov approximation for minimally coupled scalar fields.
\end{abstract}


\maketitle

\section{Introduction}
The theory of general relativity (GR) has provided us with profound insights about the fate of gravitational collapse, including the very notion of black holes \cite{Schwarzschild1916,Penrose1965}, and about the dynamics of our Universe through cosmological solutions \cite{Lemaitre1997}. It is however generally accepted that GR fails to provide an adequate description near spacetime singularities. 
In this regard, it is largely expected that the incorporation of quantum physics will provide the missing pieces to the gravitational puzzle (see e.g.~\cite{Mathur2005,Bojowald:2007ky,Ashtekaretal2018}).
Nonetheless, a definite theory of quantum gravity is yet to be discovered~\cite{Carlip2008}.

Probably, the best motivated and most conservative framework beyond GR is semiclassical gravity, both conceptually and historically~\cite{BirrellDavies1982}. In semiclassical gravity, the spacetime geometry remains classical, while the matter sector of the theory is quantized. Then, one considers the mean-value backreaction of this quantum matter onto the background geometry. This typically comes by incorporating a semiclassical Renormalized Stress-Energy Tensor (RSET) as a new source in the Einstein equations:
\begin{equation}\label{Eq:SemiEinstein}
    G_{\mu\nu}=8\pi \left( T^{\rm C}_{\mu\nu}+ \hbar\langle\hat{T}_{\mu\nu}\rangle \right).
\end{equation}
The right-hand side of this expression includes a classical SET plus the RSET, which is often represented as the vacuum expectation value of the SET operator for some quantum fields. This quantity depends on the selection of a vacuum state for the quantum fields and in general is a complicated functional (with a numeric part) of the metric functions.
The RSET can violate the pointwise energy conditions \cite{Fewster2012,Curiel2017}, thus having strong implications for the fate of horizons \cite{Zilbermanetal2022,Barceloetal2022}, gravitational collapse \cite{Barceloetal2016,AbediAfshordi2016}, and the Big Bang singularity \cite{ParkerSimon1993,Bolokhovetal2018}.

Obtaining a closed form expression for the RSET of a quantum field in specific 3+1 background spacetimes is a formidable task, even for spacetimes displaying a high degree of symmetry like static black-hole spacetimes \cite{Christiensen1976,Andersonetal1995}. Nonetheless, there have been impressive advances with the pragmatic mode sum regularization~\cite{LeviOri2016} and extended coordinate methods~\cite{Tayloretal2022}. The problem gets worse when seeking for RSETs on generic geometries, which is necessary in order to construct systems of semiclassical Einstein equations to search for their self-consistent solutions. This is the principal reason behind considering several simplifying assumptions that can make this endeavor tractable. For instance, in this paper we will restrict our discussions to a single massless quantum scalar field living in spherically-symmetric backgrounds.
In general, finding the RSET in this context requires computing accurately the modes in which the field is decomposed \cite{CandelasHoward1984}. Even for static and spherically symmetric spacetimes, the RSET can only be calculated numerically, so that performing self-consistent analyses becomes increasingly laborious. An alternative standard procedure is to resort to analytical approximations to the RSET. A suitable approximation should resemble the exact (numerical) RSET at least qualitatively, and share its main features.

\noindent
{\bf The Polyakov approximation:}
The approximation scheme that has been most extensively used in the literature is the so-called Polyakov approximation~ \cite{ParentaniPiran1994,Chakraborty2015,Carballo-Rubio2018}.
It models the RSET of a massless scalar field via a dimensional reduction to $1+1$ dimensions, where the scalar theory becomes conformally invariant. Physically, it amounts to taking only the $s$-wave sector and neglecting any back-scattering~\cite{FabbriNavarro-Salas2005}. The RSET obtained this way has the desirable properties of being analytic, second order, and unique.

The Polyakov approximation has an unphysical divergence at the radial origin, though, a limitation that comes from the very procedure used to construct this approximation. This makes it unsuitable for finding semiclassical stellar configurations. However, it is possible to regularize the Polyakov RSET using different criteria. In previous works by the authors, we analyzed in an exhaustive manner the set of semiclassical solutions that appear using a regularized Polyakov approximation. We first analyzed the vacuum solutions~\cite{Arrecheaetal2020} as well as the electrovacuum solutions~\cite{Arrecheaetal2021} using a simple cut-off regularized version. 
We found that the counterparts of the Schwarzschild and sub-extremal Reissner-Nordström black holes are non-symmetric wormholes with no horizon and an internal null singularity at a finite proper radial distance. One of the aims of the present work consists in comparing these solutions with the ones that will be obtained here using a different approximation scheme.

Later on, we analyzed the semiclassical solutions for a classical fluid of constant density~\cite{Arrechea2021b}. We found that there exist entire families of regularizations leading to self-consistent relativistic stars that are much more compact than the classical Buchdahl limit~\cite{Arrecheaetal2022}. This is possible because the semiclassical energy density becomes negative inside these stars, in a manner that generates internal layers in which the total energy density increases outward (thus violating one of the central conditions of Buchdahl theorem~\cite{Buchdahl1959}). This shows that semiclassical gravity is compatible with the existence of horizonless ultracompact objects that could mimic many of the observational properties of black holes~\cite{Carballo-Rubioetal2018,CardosoPani2019}. 

\noindent
{\bf Order-reduced AHS approximation:}
In this paper we will explore a different analytic approximation to the RSET based on applying a differential-order-reduction procedure to the Anderson, Hiscock, and Samuel (AHS) RSET~\cite{Andersonetal1995}. We will work out this procedure in detail and use it to find vacuum semiclassical solutions. Apart from developing this alternative approximation scheme, the purpose and central interest of our explorations in here is not so much to find the most precise vacuum semiclassical solution, but rather to find robust physical conclusions that are preserved when changing the approximation scheme. An example of such robust conclusions could be that in static situations event horizons disappear and get replaced by pathological singularities, as indicated by previous works~\cite{Fabbrietal2006,Berthiereetal2017,Arrecheaetal2020}.

In~\cite{Andersonetal1995}, Anderson, Hiscock and Samuel derived the RSET of a scalar field of arbitrary mass and coupling in static and spherically symmetric spacetimes, i.e. a field obeying the equation of motion
\begin{equation}\label{Eq:WaveEquation}
    \Box\phi-\left(m^2+\xi R\right)\phi=0,
\end{equation}
where $\Box$ is the d'Alembertian operator, $m$ is the field mass, and $\xi$ the coupling to the Ricci scalar $R$.
This exact RSET can be decomposed into two independently conserved parts, one of which is entirely analytical and the other requires numerical calculation,
\begin{equation}\label{Eq:AHSRSET}
    \langle\hat{T}^{\mu}_{\nu}\rangle_{\text{ren}}=\langle\hat{T}^{\mu}_{\nu}\rangle_{\text{AHS}}+\langle\hat{T}^{\mu}_{\nu}\rangle_{\text{num}}.
\end{equation}
The analytical portion $\langle\hat{T}_{\mu\nu}\rangle_{\text{AHS}}$ already captures the defining features of the standard vacuum states at horizons, yields the correct trace anomaly in the conformally invariant case, and is well defined at the center of stellar spacetimes (with a caveat shown in Appendix~\ref{Appendix:AHSReg}, where the explicit form of the AHS components is provided). This last characteristic is lacking from other RSET approximations like Polyakov's. Nonetheless, as a drawback, the analytical part (or AHS-RSET in what follows) exhibits terms that have up to fourth order derivatives of the metric functions (these expressions were also derived by Popov for the Boulware state~\cite{Popov2003}). Their presence is a consequence of the quasi-local nature of the renormalization procedure and they bring an enlargement of the space of solutions of semiclassical gravity with respect to standard GR (a straightforward demonstration of which is the enlargement of the initial conditions necessary to have a well-posed initial value problem). 

However, as it is clearly exposed in~\cite{Simon1991,ParkerSimon1993,FlanaganWald1996}, many of these solutions cannot be regarded as physical.
On the one hand, these higher-derivative equations exhibit non-physical solutions in a manner analogous to the Dirac-Abraham-Lorentz equation of classical electrodynamics~\cite{Rohrlich2000} (e.g. runaway solutions, pre-acceleration effects). Bear in mind that, while such pathologies are linked to the presence of higher-order temporal derivatives~\cite{Glavan2017}, higher-order spatial derivatives can also entail the presence of non-physical solutions, see e.g. \cite{Hochbergetal1997}.

On the other hand, there is the argument made by Simon~\cite{Simon1991} that such solutions are non-perturbative in $\hbar$, and therefore inconsistent with the idea that the semiclassical equations \eqref{Eq:SemiEinstein} are derived upon truncating the effective action of the theory at linear order in $\hbar$ \cite{HorowitzWald1980,Simon1991}. This process should eliminate all non-perturbative behaviour in $\hbar$ from the start. 
In addition to these problems, it is the case that finding the self-consistent solutions to the higher-derivative semiclassical equations by brute force, despite being doable in some situations (see \cite{Hochbergetal1997}), requires exploring an enormous space of parameters and defining a trustworthy criterion for selecting which solutions are physical. 
Henceforth, it is reasonable to subject the AHS-RSET to a procedure of order reduction so that we are left with a second-order system of differential equations.

We will work out in detail the method of reduction of order in the next section but, before that, let us advance a few observations. 
As we will see, the procedure of the order reduction naturally leads to a new analytic approximation for the RSET which satisfies all the desirable properties than an RSET should have (as captured by Wald's axioms~\cite{Wald1977}, which exact RSETs obtained through covariant renormalization procedures satisfy). We then apply our method towards obtaining geometries incorporating semiclassical backreaction in absence of classical matter. In cosmological scenarios, the order-reduced equations admit analytic solutions
\cite{Simon1992,ParkerSimon1993,Halpern1993}. On the contrary, the geometries here obtained are found through numerical integration, but analytic arguments allow to constraint the form of the various solutions that appear. 

Before ending this introduction, it is worth highlighting the advantages and limitations of the analytic approximation to the RSET obtained in~\cite{Andersonetal1995}. To which extent the very AHS-RSET is a good approximation to the total RSET depends on the background geometry on which it is evaluated and on the mass of the scalar field. Dependence on the background requires computing $\langle\hat{T}^{\mu}_{\nu}\rangle_{\text{num}}$ in each scenario and check whether the correction it entails  can be safely neglected (compared to $\langle\hat{T}^{\mu}_{\nu}\rangle_{\rm AHS}$). For conformally invariant fields in the Schwarzschild black hole spacetime, $\langle\hat{T}^{\mu}_{\nu}\rangle_{\rm AHS}$ reduces to the Page-Brown-Ottewill RSET \cite{Brownetal1986}, which has been shown to be an extremely good approximation \cite{Howard1984}. In Reissner-Nordström spacetimes, however, this approximation becomes progressively worse as the charge-to-mass ratio is increased \cite{Andersonetal1995}. Whether something similar occurs in other spacetimes (like regular stellar geometries) is yet to be known.

As for the effect of the field mass, in flat spacetime the components of the AHS-RSET reduce to
\begin{align}\label{Eq:AHSMink}
    \langle\hat{T}^{r}_{r}\rangle_{\rm AHS}=
    &
    \frac{\kappa^{4}}{1440\pi^{2}}-\frac{\kappa^{2}m^{2}}{96\pi^{2}}
    +\frac{m^{4}}{128\pi^{2}}\left(4\log\nu-3\right),\nonumber\\
    \langle\hat{T}^{t}_{t}\rangle_{\rm AHS}=
    &
    -\frac{\kappa^{4}}{480\pi^{2}}+\frac{\kappa^{2}m^{2}}{96\pi^{2}}
    +\frac{m^{4}}{128\pi^{2}}\left(4\log\nu+1\right),
\end{align}
with $\langle\hat{T}^{\theta}_{\theta}\rangle_{\rm AHS}=\langle\hat{T}^{\varphi}_{\varphi}\rangle_{\rm AHS}=\langle\hat{T}^{r}_{r}\rangle_{\rm AHS}$.
The first term on the right-hand side of~\eqref{Eq:AHSMink} is the thermal bath seen by a static observer for which the field is in a state with nonzero temperature $\kappa/2\pi$. For massive fields, $\nu$ can either equal $m e^{\gamma}/2\kappa$ or $m/2\lambda$  depending on whether the field is in a thermal or zero-temperature state, respectively ($\lambda$ is a positive parameter related to an infrared cutoff in the otherwise divergent frequency integrals in~\cite{Andersonetal1995} and $\gamma$ is Euler's constant). For massless fields, $\nu$ is an arbitrary parameter. 
Ambiguities present in RSET definitions have to be ultimately fixed via experiments. Suitable RSETs must reproduce standard results in Minkowski spacetime, where the RSET can be renormalized to zero via normal ordering~\cite{Wald1994}. In the zero-temperature massive case, the AHS-RSET~\eqref{Eq:AHSMink} has mass-dependent anomalous contributions in flat spacetime that cannot be cancelled in all components by a particular choice of $\nu$. These contributions also arise in the context of the cosmological constant problem~\cite{Martin2012}, where massive scalar fields in Minkowski generate non-vanishing quantum pressures and densities. As the involved integrals over the frequency are divergent, some regularization method must be applied. Those which violate Lorentz invariance (like a bare cutoff in the frequency) yield an effective quantum fluid that does not satisfy the equation of state of vacuum energy. The infrared cutoff $\lambda$ plays a similar role here, as the components~\eqref{Eq:AHSMink} describe a cosmological constant only in the $\nu\to0$ and $\nu\to\infty$ limits, the latter corresponding to the absence of any infrared cutoff $\lambda$ in the renormalization procedure. In that case,  the RSET components~\eqref{Eq:AHSMink} diverge, although in a way that satisfies the equation of state of vacuum energy.
As this discussion extends beyond the scope of this article, we adopt the view that the analytic approximation alone cannot be trusted if the field is massive since, in black hole spacetimes, their components will be non-vanishing in the asymptotically flat region for the Boulware state. Consequently, we set $m=0$ for the remaining of this work.
 
The paper is organised as follows. Section \ref{Sec:Order} describes the order-reduction procedure and the derivation of the order-reduced RSET approximation in the Boulware vacuum state. Then, in section \ref{Sec:Solutions} we solve the semiclassical equations self-consistently using this new RSET approximation and analyze the characteristics of the solutions. Section \ref{Sec:Validity} discusses the accuracy of the order-reduced RSET by addressing how different is this OR-RSET from the AHS-RSET for particular solutions. We compare these results with previous analyses using the regularized Polyakov approximation in section \ref{Sec:Comparison}. In particular, we discuss which elements are specific of the approximation and which ones are more generic. We conclude with a summary of our results and some final remarks.

\section{Reducing the order of the AHS-RSET in vacuum}\label{Sec:Order}

The order reduction procedure did not originate in semiclassical analyses, but is much older and emerged in the context of the electromagnetic radiation reaction equation~\cite{Landau1975}. The method applies to any theory where higher-order contributions in a set of ordinary differential equations are multiplied by some small parameter~\cite{Beletal1981,BelZia1985} in terms of which the solutions can be expanded.

In the context of semiclassical gravity, order reduction was first applied to prove the stability of flat space~\cite{Simon1991} and the absence of a Starobinski inflationary phase driven by semiclassical backreaction~\cite{Simon1992}. Then, it has been applied as well for obtaining the dynamics of FLRW and Kasner cosmologies \cite{ParkerSimon1993,SiemieniecOzieblo1999,Halpern1993} and the study of averaged energy conditions \cite{FlanaganWald1996}. Here, we apply this method to static, spherically symmetric spacetimes, the line element of which can be written, without loss of generality, as
\begin{equation}\label{Eq:LineElement}
    ds^{2}=-f(r)dt^{2}+h(r)dr^{2}+r^{2}d\Omega^{2},
\end{equation}
where $d\Omega^{2}$ is the line element of the unit $2$-sphere.

Not all components of the vacuum semiclassical Einstein equations [i.e. Eq. \eqref{Eq:SemiEinstein} without the classical SET] are independent. In fact, we can just focus on the $tt$ and $rr$ components:
\begin{align}\label{Eq:SemiComps}
     \frac{h(1-h)-rh'}{h^{2}r^{2}}=
    &
    ~8\pi\hbar\langle \hat{T}^{t}_{t}\rangle_{\text{AHS}},\nonumber\\
    \frac{rf'f-fh}{fhr^{2}}=
    &
    ~8\pi\hbar\langle \hat{T}^{r}_{r}\rangle_{\text{AHS}},
\end{align}
where the right-hand side contains higher-derivative terms. As mentioned above, this RSET was found by Anderson, Hiscock and Samuel. Its concrete and lengthy form, which is not very illustrative, can be seen in~\cite{Andersonetal1995}; we recall it here in Appendix~\ref{Appendix:AHS} for completeness.
Neglecting terms $\order{\hbar}$ in~\eqref{Eq:SemiComps} leads to
\begin{align}\label{Eq:hbarExp}
    \frac{h(1-h)-rh'}{h^{2}r^{2}}=
    &
    ~\order{\hbar},\nonumber\\
    \frac{rf'+f-fh}{fhr^{2}}=
    &
    ~\order{\hbar}.
\end{align}
These expressions can be differentiated consecutively to derive recursion relations between $f,~h$, and their higher-order derivatives $\{f^{(n)}\}_{n=1}^\infty$ and $\{h^{(n)}\}_{n=1}^\infty$. For $h$, said relations are obtained by solving the $tt$ equation directly, which can then be used to derive the $f$ relations from the $rr$ equation. The result is
\begin{align}\label{Eq:OrderRels}
    h^{(n)}=
    &
    \left(-1\right)^{n}\frac{n!h^{n}}{r^{n}}\left(h-1\right)+\order{\hbar},\nonumber\\
    f^{(n)}=
    &
    \left(-1\right)^{n+1}\frac{n!f}{r^{n}}\left(h-1\right)+\order{\hbar}.
\end{align}

Relations \eqref{Eq:OrderRels} are now inserted in the AHS-RSET components $\langle\hat{T}^{t}_{t}\rangle_{\text{AHS}}$ and $\langle\hat{T}^{r}_{r}\rangle_{\text{AHS}}$ 
[Eqs. \eqref{Eq:AHSRSETtt} and \eqref{Eq:AHSRSETrr} in Appendix~\ref{Appendix:AHS}]
until they only depend on $f$ and $h$.
After a lengthy but straightforward calculation using symbolic computation software, the resulting expressions are

\begin{align}\label{Eq:VOR}
    16\pi^{2}\langle\hat{T}^{t}_{t}\rangle_{\rm OR}=
    &
    -\frac{\kappa ^4}{30 f^2}+\kappa^{2}\left(\xi -\frac{1}{6}\right)\frac{3\left(h-1\right)^{2} }{6f h r^2}\nonumber\\
    &
    +\frac{\left(h-1\right)^{2}\left(h^{2}+6h+33\right)}{480h^{2}r^{4}}\nonumber\\
    &
    -\left(\xi -\frac{1}{6}\right)\frac{\left(h-1\right)^{2}\left(h^{2}+2h+5\right)}{8h^{2}r^{4}},\nonumber\\
    16\pi^{2}\langle\hat{T}^{r}_{r}\rangle_{\rm OR}=
    &
    ~\frac{\kappa ^4}{90 f^2}-\kappa^{2} \left(\xi -\frac{1}{6}\right)\frac{\left(h-1\right)\left(h+3\right)}{ 6fh r^2}\nonumber\\
    &
    -\frac{\left(h-1\right)^{2}\left(h^{2}+6h-15\right)}{1440h^{2}r^{4}} \nonumber\\
    &
    +\left(\xi-\frac{1}{6}\right)\frac{\left(h-1\right)^{2}\left(h+3\right)^{2}}{24h^{2}r^{4}},
\end{align}
where the suffix OR stands for order-reduced.
Taking $\kappa=0$ selects the Boulware vacuum state; this is the state considered here as it is the one consistent with staticity and asymptotic flatness. Being a zero-temperature state for static observers, the RSET in the Boulware state is divergent at the event horizon~\cite{Boulware1974}. On the other hand, by taking $\kappa$ equal to the surface gravity of the Schwarzschild black hole we ensure the finiteness of the RSET components at the horizon, thus selecting the Hartle-Hawking vacuum state.

So far, we have applied the method of order reduction to the $tt$ and $rr$ components of the RSET. 
If we continue and apply the same method to the angular components of the RSET, it turns out that the set of 
order-reduced field equations do not satisfy the Bianchi identities~\cite{BelZia1985}. In other words, we find that the order-reduced RSET is not covariantly conserved, but satisfies
\begin{equation}\label{Eq:ConsVOR}
    \nabla_{\mu}\langle\hat{T}^{\mu}_{\nu}\rangle_{\rm OR}=\order{\hbar}.
\end{equation}
However, as it is discussed just below, we propose a different algorithm for order reduction that leads to a  
covariantly conserved RSET. This will be used in our analysis of solutions.

\subsection{The covariantly conserved OR-RSET and its properties}

There is a straightforward way to obtain a covariantly conserved RSET. The idea is simple: once we have reduced the order of the $tt$ and $rr$ components of the AHS-RSET, we add the angular components needed to obtain conservation. This is an unambiguous procedure under the assumptions of staticity and spherical symmetry. In fact, the only non-trivial component of the divergence of the OR-RSET \eqref{Eq:ConsVOR} is 
\begin{align}
    \nabla_{\mu}\langle\hat{T}^{\mu}_{r}\rangle_{\rm{OR}}=
    &
    ~\partial_{r}\langle\hat{T}^{r}_{r}\rangle_{\rm{OR}}+\frac{2}{r}\left(\langle\hat{T}^{r}_{r}\rangle_{\rm{OR}}-\langle\hat{T}^{\theta}_{\theta}\rangle_{\rm{OR}}\right)\nonumber\\
    &
    +\frac{f'}{2f}\left(\langle\hat{T}^{r}_{r}\rangle_{\rm{OR}}-\langle\hat{T}^{t}_{t}\rangle_{\rm{OR}}\right).
\end{align}
Using the expressions for $\langle\hat{T}^{t}_{t}\rangle_{\rm{OR}}$ and $\langle\hat{T}^{r}_{r}\rangle_{\rm{OR}}$ in Eq.~\eqref{Eq:VOR} we can deduce the angular components necessary to force this divergence to vanish. Specifically, we find 
\begin{widetext}
\begin{align}\label{Eq:Tthth}
16\pi^{2}\langle\hat{T}^{\theta}_{\theta}\rangle_{\rm{OR}}=
16\pi^{2}\langle\hat{T}^{\varphi}_{\varphi}\rangle_{\rm{OR}}=
&
~\frac{\kappa ^4}{90 f^2}-\kappa^{2}\left(\xi-\frac{1}{6}\right)\frac{f\left(h^{2}+3\right)+f'\left(h-1\right)\left(h-3\right)}{12f^{2}hr}\nonumber\\
&
\hspace{-2cm}-\frac{h-1}{1440fh^{2}r^{4}}\left\{f\left[h^{3}\left(r-1\right)+h^{2}\left(3r-5\right)+3h\left(r+7\right)-15\left(r+1\right)\right]+f'\left(h-1\right)\left(h^{2}+6h+21\right)\right\}\nonumber\\
&
\hspace{-2cm}+\left(\xi-\frac{1}{6}\right)\frac{h-1}{24fh^{2}r^{4}}\left\{f\left(h+3\right)\left[h^{2}\left(r-1\right)-2h+3\left(r+1\right)\right]+rf'\left(h-1\right)\left(h^{2}+3h+6\right)\right\}.
\end{align}
\end{widetext}
We have thus constructed an approximation to the RSET valid for the study of vacuum spacetimes. 
The OR-RSET approximation is divergenceless, yields the correct values at the event horizon, and vanishes in flat spacetime in the Boulware state. It therefore satisfies all Wald's axioms~\cite{Wald1977}.
In addition, it is not a higher-derivative quantity, which makes it an excellent RSET candidate to address the semiclassical backreaction problem. In principle, it is possible to select $\langle\hat{T}^{r}_{r}\rangle_{\rm AHS}$ and $\langle\hat{T}^{\theta}_{\theta}\rangle_{\rm AHS}$ as the components to reduce, and determine $\langle\hat{T}^{t}_{t}\rangle$ via Eq.~\eqref{Eq:ConsVOR}. This results in an RSET different from the OR-RSET and whose $\langle\hat{T}^{t}_{t}\rangle$ component is of greater derivative order than~\eqref{Eq:VOR}. The procedure followed here yields the simplest conserved, low-order RSET among all possible choices of order-reduction scheme.

Before turning to analyze the set of self-consistent solutions of these semiclassical equations, let us make a few further observations. On the one hand, the reduction of order in the vacuum case gets rid of the $\nu$ parameter. This allows us to give unique and unambiguous results for the corresponding solutions, and simplifies the analysis significantly.

On the other hand, when the OR-RSET is evaluated on the Schwarzschild spacetime, $f(r)=h(r)^{-1}=1-2M/r$, we realize that it gives the exact result  
\begin{equation}\label{Eq:SchwEq}
\langle\hat{T}^{\mu}_{\nu}\rangle_{\rm{OR}}^{\text{(Schw)}}=\langle\hat{T}^{\mu}_{\nu}\rangle_{\text{AHS}}^{\text{(Schw)}}.    
\end{equation}
This is not surprising upon realizing that the Schwarzschild metric is an exact solution to Eq.~\eqref{Eq:hbarExp}, thus the $\order{\hbar^{2}}$ terms neglected in our expansion vanish identically. Therefore, for the Schwarzschild metric only, the reduction of order is an exact procedure, i.e. the OR-RSET coincides with the AHS-RSET. The AHS-RSET also returns the correct trace anomaly for conformally invariant fields ($m=0,~\xi=1/6$). The trace of the OR-RSET differs from the trace anomaly but it is manifestly state-independent in the case of conformal coupling. 

Even though in this paper we are not going to consider the non-vacuum case, let us advance here that, in that case, the order-reduction method shows some differences.
In particular, the analogue of Eqs.~\eqref{Eq:OrderRels} acquire extra contributions proportional to the classical stress-energy tensor
$T^{\rm C}_{\mu\nu}$. Nonetheless, using the conservation relation [analogous to~\eqref{Eq:ConsVOR}] and some equation of state for the classical SET we find an order-reduced RSET that can be decomposed as
\begin{equation}\label{Eq:MOR}
    \langle\hat{T}^{\mu}_{\nu}\rangle_{\rm{MOR}}=\langle\hat{T}^{\mu}_{\nu}\rangle_{\rm{OR}}+\langle\hat{\mathcal{F}}^{\mu}_{\nu}\rangle+\hat{\mathcal{G}}^{\mu}_{\nu}\log{\nu},
\end{equation}
where MOR stands for matter-order-reduced. Here, $\langle\hat{\mathcal{F}}^{\mu}_{\nu}\rangle$ (brackets indicate that this term is $\kappa$-dependent) and $\hat{\mathcal{G}}^{\mu}_{\nu}$ are functions of the classical SET and vanish whenever the latter vanishes, recovering the OR-RSET in vacuum. Every term in Eq.~\eqref{Eq:MOR} is conserved independently, since $\langle\hat{T}^{\mu}_{\nu}\rangle_{\rm{MOR}}$ is conserved by construction and for any $\nu$.
Finally, the MOR-RSET is regular in any spacetime that does not exhibit curvature singularities at $r=0$, solving the caveat present in the AHS-RSET and detailed in Appendix~\ref{Appendix:AHSReg}. This puts the MOR-RSET on equal footing with regularized versions of the Polyakov approximation.


\section{Classification of vacuum solutions}\label{Sec:Solutions}

Equipped with the OR-RSET from Eqs.~(\ref{Eq:VOR}) and~(\ref{Eq:Tthth}), we are now prepared to address the backreaction problem. In principle, the domain of consistency of the order reduction \textit{à la} Simon~\cite{Simon1991} is limited to solving the order-reduced equations (expressions displayed below) perturbatively~\cite{FlanaganWald1996}. This requires assuming from the onset that the Einstein tensor and the OR-RSET admit perturbative expansions in $\hbar$ and then solve the expanded equations order by order. Whereas entirely consistent, this logic eliminates the possibility of any non-perturbative backreaction effect associated with the Boulware state (recall that this state is singular at event horizons), which any proper RSET should capture as well. We adopt a modified gravity perspective hereafter and solve the semiclassical equations in a self-consistent fashion, that is, determining the spacetime geometry and the sources that generate it simultaneously.
We will elaborate on the logic behind this approach and on the comparison between results derived with different RSET approximations in Sec.~\ref{Sec:Comparison}. For the moment, it just necessary to keep in mind that, regardless of its origin, the OR-RSET satisfies all the requirements for a legitimate semiclassical source, and in the following we are treating it as such.

In this section, we will classify the self-consistent semiclassical solutions in vacuum for arbitrary coupling $\xi$ and ADM mass $M$. The geometries here depicted also describe, for $M>0$, the spacetime exterior to any static stellar configuration in this approximation.

The $tt$ and $rr$ components of the semiclassical equations with the OR-RSET in Eq.~\eqref{Eq:VOR} can be cast in the form of a dynamical system
\begin{align}\label{Eq:hFieldEq}
    h'=
     &
     -\frac{\left(h-1\right)h}{r}+\frac{l_{\rm P}^{2}}{r^{3}}\left(h-1\right)^{2}\left[\left(\xi-\xi_{\rm c}\right)h^{2}\vphantom{\left(\xi-\frac{83}{300}\right)}\right.\nonumber\\
     &
     \left.+2\left(\xi-\xi_{\rm c}+\frac{1}{30}\right)h+5\left(\xi-\frac{83}{300}\right)\right],\\
    \frac{f'}{f}=
    &
    ~\frac{\left(h-1\right)}{r}+\frac{l_{\rm P}^{2}}{3hr^{3}}\left(h-1\right)^{2}\left[\left(\xi-\xi_{\rm c}\right)h^{2}\vphantom{\left(\xi-\frac{25}{3}\right)}\right.\nonumber\\
    &
    \left.+6\left(\xi-\xi_{\rm c}\right)h+9\left(\xi-\frac{25}{3}\right)\right],\label{Eq:fFieldEq}
\end{align}
where $l_{\rm P}^{2}=\hbar/16\pi$ and $\xi_{\rm c}=11/60$. We have solved this system of equations numerically by imposing asymptotically flat boundary conditions at some distant radius, namely requiring that
\begin{equation}\label{Eq:AsymptConds}
    f\left(r\right)=h\left(r\right)^{-1}=1-\frac{2M}{r},
\end{equation}
in the limit $r\to\infty$ (in practice, we are considering $r\gg M$).
Depending on the sign of the ADM mass $M$, we find two distinct types of solutions. The most interesting regime for us is $M>0$, which corresponds to the Schwarzschild black hole in the classical theory. Furthermore, for \mbox{$M>0$}, there is a critical value of the coupling 
\begin{equation}\label{Eq:CritCoup}
   \xi_{\rm c} = 11/60,
\end{equation}
that denotes a separatrix between two regimes of solutions. The special value $\xi_{\rm c}$ appears as a correction (originated by the trace anomaly) to the conformal coupling $\xi = 1/6$ for which Eq.~\eqref{Eq:WaveEquation} becomes invariant under conformal transformations.


In what follows, we turn to describe the main features of the numerical solutions we have obtained, while also providing analytical arguments to constrain the form of the solutions whenever possible.

\subsection{Positive asymptotic mass ($M>0$)}

\subsubsection{Coupling $\xi<\xi_{\rm c}$}
Starting from the asymptotic expressions~\eqref{Eq:AsymptConds}, it is possible to constrain the form of the solutions to Eqs.~(\ref{Eq:hFieldEq},~\ref{Eq:fFieldEq}) for the case $\xi<\xi_{\rm c}$ through simple analytic arguments. First, we turn our attention to Eq.~\eqref{Eq:hFieldEq}. It is straightforward to check that its right-hand side is always negative. Conditions~\eqref{Eq:AsymptConds} guarantee that $h>1$ asymptotically, so $h$ increases inwards with no turning points.
In principle, the function $h$ could adopt one of the following behaviours: i) reaching a finite value at $r=0$, ii) diverging at $r=0$, or iii) diverging at $r>0$. As for possibility i), the semiclassical equations
enforce the conditions $h(0)=1$ and $h'(0)=0$. Let us prove this statement by assuming that the following expansions hold for the metric functions around $r=0$:
\begin{equation}\label{Eq:ExpsOr}
    h=\sum_{n=0}^{\infty}h_{n}r^{n},\quad f=\sum_{n=0}^{\infty}f_{n}r^{n}.
\end{equation}
By replacing them in 
Eqs.~(\ref{Eq:hFieldEq},~\ref{Eq:fFieldEq}), we obtain the following values for the lowest order expansion parameters
\begin{align}\label{Eq:Coefs}
    h_{0}=
    &
    1,\quad h_{1}=f_{1}=h_{3}=f_{3}=0,\nonumber\\
    h_{2}=
    &
    \frac{3}{2l_{\rm P}^{2}\left(4\xi-1\right)}, \quad
    f_{2}=
    \frac{\left(90\xi-17\right)}{15\left(4\xi-1\right)}f_{0},
\end{align}
with $f_{0}>0$ a constant denoting the value of the redshift at the origin. Under these conditions, $r=0$ is a regular point of the solution. Now, since we have $h>1$ asymptotically and $h$ cannot have turning points, possibility i) is discarded, as a finite value of $h$ at $r=0$ must be a maximum (or minimum). Regarding the behaviour ii), an infinite $h(0)$ is inconsistent with Eq.~\eqref{Eq:hFieldEq} which, at leading order, would take the form
\begin{equation}
    h'\propto\frac{h^{4}}{r^{3}}.
\end{equation}
Solutions are inconsistent with assuming a divergent $h$ at $r=0$. Thus, we must discard possibility ii) as well. In conclusion, iii) holds and $h$ must diverge at some radius $r=r_{\rm D}>0$.

Let us highlight that the OR-RSET is finite for spherically symmetric spacetimes which are regular at $r=0$. This is so even under the fact that the OR-RSET is derived assuming the absence of any classical matter, for which the resulting classical solutions are all singular (excluding flat spacetime, for which the RSET vanishes in the Boulware state).
This characteristic is appealing because it suggests that our method is well defined for stellar spacetimes, a characteristic absent from the Polyakov~\cite{Polyakov1981} and $s$-wave approximations~\cite{FabbriNavarro-Salas2005}. This regularity is essential for the construction of complete semiclassical stellar models. See Appendix \ref{Appendix:AHSReg} for a discussion on the regularity of the AHS-RSET.


We now turn our attention to the differential equation for $f$, Eq.~\eqref{Eq:fFieldEq}. In virtue of \eqref{Eq:AsymptConds} we have, asymptotically, $f>0$ and $f'>0$. Again, the term between brackets on the right-hand side of the equation is negative. Thus, if $h$ increases monotonically as $r$ decreases, these negative terms will eventually compensate the positive contribution from the first term on the right-hand side, generating a turning point in the $f$ function.

We analyze now the divergent behaviour of $h$ at some radius $r=r_{\rm{D}}>0$. Close to $r_{\rm D}$, Eqs. (\ref{Eq:hFieldEq},~\ref{Eq:fFieldEq}) are approximated by
\begin{align}
    h'\simeq
    &
    \frac{l_{\rm P}^{2}}{r_{\rm{D}}^{3}}\left(\xi-\xi_{\rm c}\right)h^{4},\nonumber\\
    \frac{f'}{f}\simeq
    &
    \frac{l_{\rm P}^{2}}{3r_{\rm{D}}^{3}}\left(\xi-\xi_{\rm c}\right)h^{3}.
\end{align}
The differential equation for $h$ can be integrated directly, yielding
\begin{equation}\label{Eq:Exph1}
    h\simeq r_{\rm{D}}\left[3l_{\rm P}^{2}\left(\xi_{\rm c}-\xi\right)x\right]^{-1/3},
\end{equation}
where $x=r-r_{\rm D}$. Replacing this expression in the equation for $f$ returns
\begin{equation}\label{Eq:Expf1}
    f\simeq\frac{f_{\rm D}}{x^{1/9}},
\end{equation}
where $f_{\rm D}$ is a positive integration constant bearing a relation with the ADM mass $M$, not relevant here.

The curvature scalar adopts the expression
\begin{align}\label{Eq:Rscalar}
    R=
    &
    \frac{2}{r^2}\left(1-\frac{1}{h}\right)+\frac{2}{hr}\left(\frac{h'}{h}-\frac{f'}{f}+\frac{rf'h'}{4fh}\right)\nonumber\\
    &
    +\frac{1}{2h}\left[\left(\frac{f'}{f}\right)^{2}-\frac{2f''}{f}\right].
\end{align}
Near $r=r_{\rm D}$, where the approximate solutions~\eqref{Eq:Exph1} and~\eqref{Eq:Expf1} hold, the curvature scalar~\eqref{Eq:Rscalar} becomes
\begin{equation}\label{Eq:Rcase1}
    R\simeq-\chi_{0}x^{-5/3},
\end{equation}
with
\begin{equation}
    \chi_{0}=\frac{8}{9 r_{\rm D}}\left[\frac{l_{\rm P}^{2}\left(\xi_{\rm c}-\xi\right)}{9}\right]^{1/3}.
\end{equation}

In view of Eq.~\eqref{Eq:Rcase1}, the metric has a curvature singularity at $r=r_{\rm{D}}$. This is a consequence of allowing the Boulware state to backreact onto the background geometry in a self-consistent (and non-perturbative) way.

\begin{figure*}
    \centering
    \includegraphics[width=\textwidth]{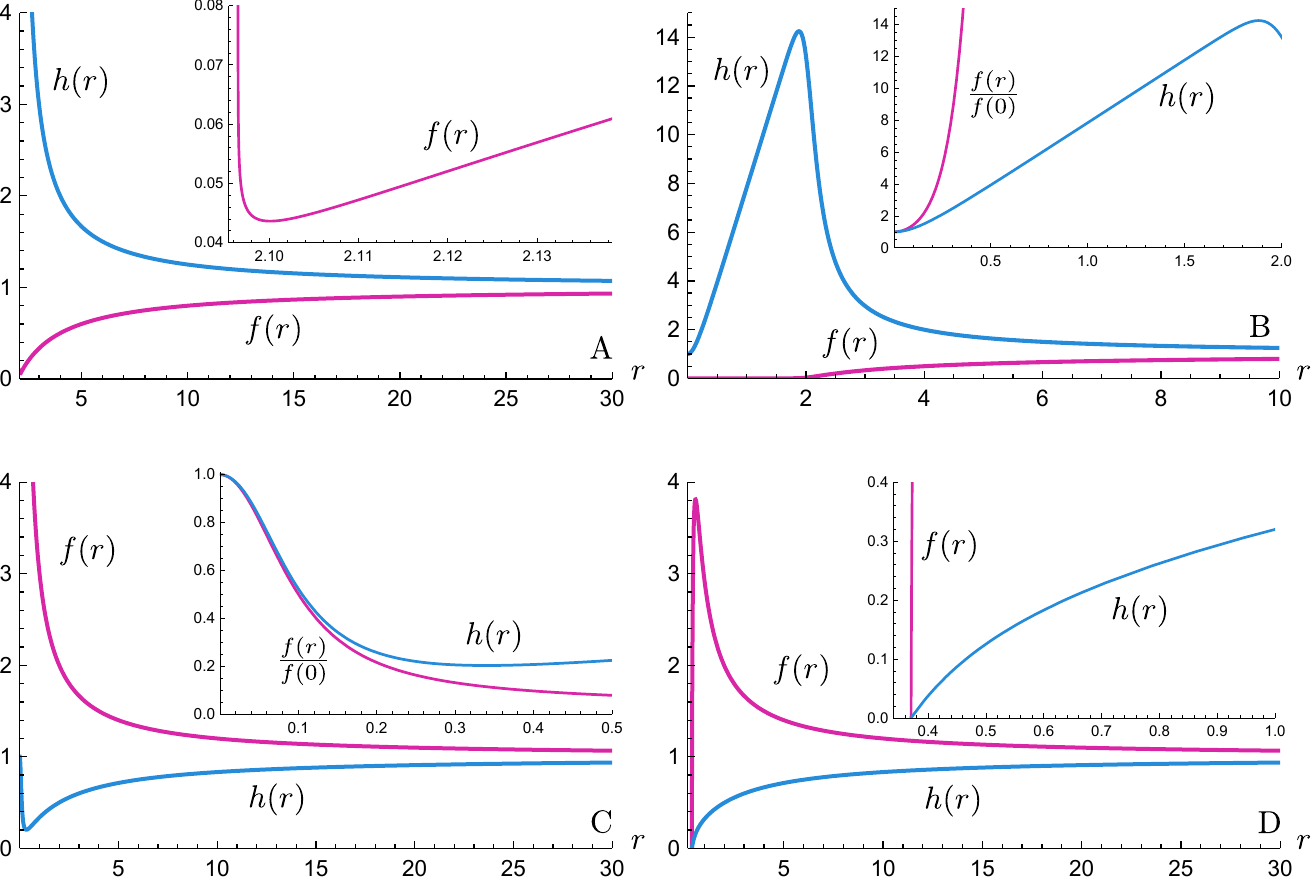}
    \caption{Panels A and B: self-consistent semiclassical solutions with $M=1$ for fields of coupling $\xi=0$ and $\xi=1$, respectively (the values have been chosen for illustrative purposes). The geometry with $\xi\leq\xi_{\rm c}$ (in particular, $\xi=0$) is a timelike naked singularity whereas for $\xi>\xi_{\rm c}$ (in particular, $\xi=1$) we have a regular geometry with Planckian curvatures. Panels C and D: self-consistent semiclassical solutions with $M=-1$ for fields of coupling $\xi=0$ and $\xi=1$, respectively. For $\xi<83/300$ (in particular, $\xi=0$), the geometry describes a core of negative Misner-Sharp mass $m(r)$, at whose center the redshift can be maximal (if $\xi\leq\xi_{\rm c}$) or minimal (if $\xi_{\rm c}<\xi\leq83/300$). For $\xi>83/300$ (in particular, $\xi=1$) the geometry has a null singularity at the surface where $h$ and $f$ vanish.}
    \label{Fig:Metrics}
\end{figure*}

\begin{figure}
    \centering
    \includegraphics[width=\columnwidth]{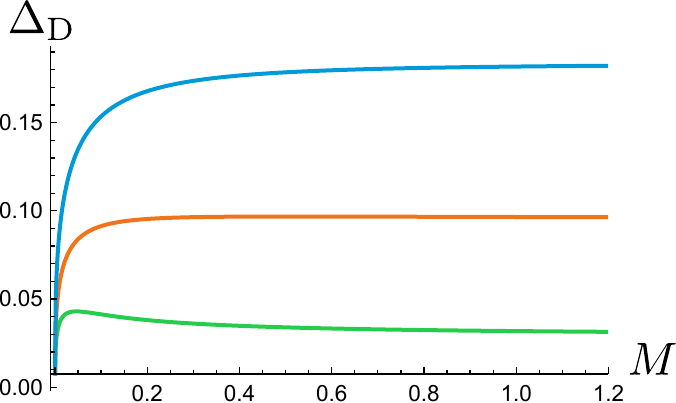}
    \caption{Radial distance $\Delta_{\rm D}=r_{\rm D}-2M$ as a function of the ADM mass $M$ for different values of the coupling $\xi$. The green, orange and blue curves correspond to \mbox{$\xi=\left\{1/6,0,-1/2\right\}$}, respectively. The distance $\Delta_{\rm D}$ vanishes in the $M\to0$ limit and quickly reaches a constant value as $M$ increases.}
    \label{Fig:Deltar}
\end{figure}

\begin{figure}
    \centering
    \includegraphics[width=\columnwidth]{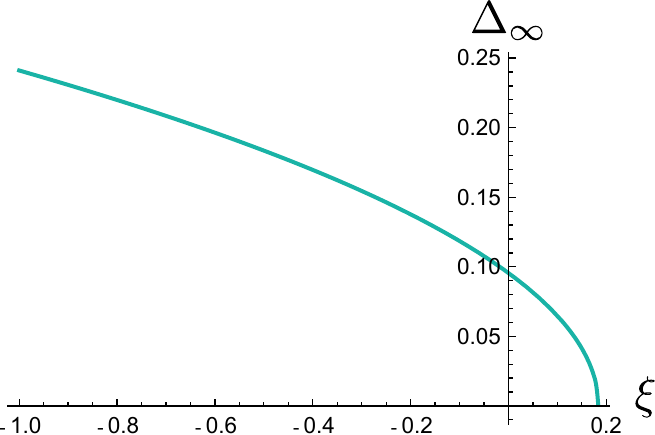}
    \caption{Asymptotic value of the radial distance $\Delta_{\infty}$ obtained as $M$ is increased. This quantity diverges as $\xi\to-\infty$ and decreases with increasing $\xi$, vanishing in the $\xi\to\xi_{\rm c}$ limit. }
    \label{Fig:DistCoupling}
\end{figure}

The reader can consult panel A in Fig. \ref{Fig:Metrics} for plots of the metric functions in terms of $r$. As numerical integrations of Eqs.~(\ref{Eq:hFieldEq},~\ref{Eq:fFieldEq}) reveal, $r_{\rm D}$ is a surface located slightly above the classical Schwarzschild radius $2M$. The radial distance \mbox{$\Delta_{\rm D}=r_{\rm D}-2M$} depends on the ADM mass of the geometry for every fixed value of the coupling $\xi$. Fig. \ref{Fig:Deltar} contains some example cases for various $\xi$. As $M$ increases, $\Delta_{\rm D}$ tends to the $\xi$-dependent constant $\Delta_{\infty}$. Figure \ref{Fig:DistCoupling} shows that $\Delta_{\infty}$ decreases as $\xi$ increases, vanishing in the $\xi\to\xi_{\rm c}$ limit, i.e. the configuration with the highest $\xi$ belonging to this family.


\subsubsection{Coupling $\xi=\xi_{\rm c}$}

Now we turn to analyze the particular case where \mbox{$\xi=\xi_{\rm c}$}. In this situation, similar analytical arguments as the ones exhibited above apply to Eq.~\eqref{Eq:hFieldEq}, that is, $h$ grows monotonically inwards until it diverges at some finite radius $r_{\rm D}$ as above. The $f$ function, however, does not reach a turning point, since the $\order{h^{2}}$ and $\order{h}$ terms between brackets at the right-hand side in~\eqref{Eq:fFieldEq} vanish and the $\order{h^{0}}$ terms are suppressed by $l_{\rm P}^{2}$ with respect to the $\order{l_{\rm P}^{0}}$ terms. Close to $r_{\rm D}$, the field equations admit the expansions
\begin{align}
    h'\simeq
    &
    -\frac{l_{\rm P}^{2}}{15r_{\rm{D}}^{3}}h^{3},\nonumber\\
    \frac{f'}{f}\simeq
    &
    \frac{\left(15r_{\rm{D}}^{2}+2l_{\rm P}^{2}\right)}{15r_{\rm{D}}^{3}}h.
\end{align}
Integrating them and expanding the solution for $f$ in powers of $x=r-r_{\rm D}$ we have
\begin{align}
    h\simeq
    &
    \sqrt{\frac{15r_{\rm{D}}^{3}}{2l_{\rm P}^{2}}}x^{-1/2},\nonumber\\
    f\simeq
    &
    f_{\rm c}\left[1+\frac{60r_{\rm{D}}^{2}+8l_{\rm P}^{2}}{\sqrt{2 r_{\rm{D}}^{3}}l_{\rm P}}
    x^{1/2}\right],
\end{align}
with $f_{\rm c}>0$. The critical solution $\xi=\xi_{\rm c}$ has divergent $h$ but vanishing redshift function $f$ at $r_{\rm D}$. This results again in a curvature singularity at $r=r_{\rm D}$ such that the curvature scalar
\begin{equation}
    R\simeq-\frac{15r_{\rm{D}}^{2}+2l_{\rm P}^{2}}{60r_{\rm{D}}}x^{-1},
\end{equation}
diverges slower than in Eq.~\eqref{Eq:Rcase1}.

\subsubsection{Coupling $\xi>\xi_{\rm c}$}
Increasing the coupling beyond its critical value $\xi_{\rm c}$ results in geometries of a drastically different nature.

We draw our attention to Eq.~\eqref{Eq:hFieldEq} first. At large $r$ there is an initial tendency of $h$ to increase inwards which will eventually be reverted by the (now positive) contribution coming from the $\order{h^{4}}$ term in the right-hand side of~\eqref{Eq:hFieldEq}. Hence, $h$ cannot diverge towards positive infinity at finite radius. Similarly, $h$ cannot vanish at some positive $r$ since that would necessarily imply $h=1$ there. Because of \eqref{Eq:hFieldEq}, $h$ is bounded from below by the value $h=1$, which corresponds to a minimum for $h$. The only remaining possibility is that $h$ goes to a constant value at $r=0$. Recall that 
the only value of $h(0)$ consistent with Eq.~\eqref{Eq:hFieldEq} is $h(0)=1$ ---see Eqs. (\ref{Eq:ExpsOr},~\ref{Eq:Coefs})--- which corresponds to a finite Ricci scalar at $r=0$. On the other hand, the $f$ function decreases monotonically from the asymptotic region inwards, reaching a finite value at $r=0$, as the term within brackets in Eq.~\eqref{Eq:fFieldEq} is everywhere positive for $\xi>\xi_{\rm c}$.

The metric functions obtained from numerical integration of the Eqs. (\ref{Eq:hFieldEq},~\ref{Eq:fFieldEq}) are displayed in Fig. \ref{Fig:Metrics}B. Despite being regular, these spacetimes are not well defined from a semiclassically-consistent perspective, as their curvature scalars have Planckian magnitudes. Indeed, inserting Eqs.~\eqref{Eq:ExpsOr} and~\eqref{Eq:Coefs} into Eq.~\eqref{Eq:Rscalar}, we obtain
\begin{equation}\label{Eq:Rcase2}
    R=\frac{45\left( \xi -16/75\right)}{8l_{\rm P}^2(\xi -1/4)^2}+\order{r^{2}},
\end{equation}
close to the origin.

The metrics from Figs. \ref{Fig:Metrics}A and \ref{Fig:Metrics}B are similar for \mbox{$r\gg2M$}, i.e. the region where semiclassical corrections are perturbative. Thus, within their regimes of validity, these geometries approximate well the exterior spacetime of any semiclassical stellar object for any field coupling $\xi$ to the curvature . We will define a criteria to estimate the closeness between solutions to the order-reduced and higher-order systems in the next section.

So far we have classified the space of vacuum solutions with positive ADM mass. For couplings $\xi\leq\xi_{\rm c}$ we find geometries where the event horizon gets replaced by a naked curvature singularity. For couplings $\xi>\xi_{\rm c}$ we obtain regular spacetimes with curvatures that become Planckian below the region where the event horizon would have appeared. This drastic change of behaviour can be traced back to which pointwise energy conditions~\cite{Visser1996} the OR-RSET violates depending on the coupling near the would-be event horizon: for $\xi\leq \xi_{\rm c}$ the null energy condition is violated and the strong energy condition holds, whereas for $\xi>\xi_{\rm c}$ both the null and strong energy conditions are violated.

\subsection{Negative asymptotic mass ($M<0$)}
For the sake of completeness, we provide here detailed descriptions of the geometries obtained when Eqs.~(\ref{Eq:hFieldEq},~\ref{Eq:fFieldEq}) are integrated for negative ADM mass. In the classical theory, $M<0$ corresponds to a naked singularity. In some situations, the semiclassical backreaction can even regularize these singularities. This characteristic is of interest to the study of stellar spacetimes for reasons that will become clear in what follows. 

We first address the case $\xi\leq\xi_{\rm c}$. For $M<0$ we have in virtue of Eq.~\eqref{Eq:AsymptConds} that $h<1$, $h'>0$, $f>1$, and $f'<0$ asymptotically. By inspection of Eq.~\eqref{Eq:hFieldEq} we observe the RSET contribution (i.e. the terms within brackets) is everywhere negative for $\xi\leq\xi_{\rm c}$. As the equations are integrated inwards, these contributions will decrease, cancelling the $\order{l_{\rm P}^{0}}$ term, until $h$ reaches a minimum. Afterwards, consistency of Eq.~\eqref{Eq:hFieldEq} indicates that $h$ can only grow inwards until reaching $h=1$ at $r=0$, where the resulting solution is described by the expansions~\eqref{Eq:Coefs} derived previously (which hold for $M<0$ as well). On the other hand, $f$ must increase monotonically inwards (since the right hand side of~\eqref{Eq:fFieldEq} is always negative) and it can only reach a finite value at $r=0$ [again obeying Eq.~\eqref{Eq:Coefs}], while $h$ lies always below $h=1$ for $r>0$, which would correspond to a maximum. 

Near $r=0$, the expansion of the metric functions~\eqref{Eq:Coefs} now enforces $h_{2}<0$ and $f_{2}>0$. Numerical integration of the semiclassical equations is displayed in Fig.~\ref{Fig:Metrics}C. This solution has a clear interpretation in terms of the Misner-Sharp mass $m(r)=r\left(1-h^{-1}\right)/2$~\cite{MisnerSharp1964,HernandezMisner1966}, which is negative everywhere.
In the classical solution, this mass would be $m(r)=-M$ and would inevitably generate a curvature singularity at $r=0$. Here, the zero-point energies from the scalar field provide a (positive) contribution to this negative mass that balances it exactly at $r=0$. The Ricci scalar is negative and Planckian at $r=0$.

The semiclassical equations in an order-reduced version accommodate solutions that display whole regions of negative mass that get regularized at $r=0$ by semiclassical effects. This is a characateristic that we observed in ultracompact stellar configurations using a regularized version of the Polyakov RSET~\cite{Arrechea2021b}. So far, these results look appealing in the sense that, if the innermost layers of a semiclassical star with a negative mass interior generate relevant semiclassical contributions, these can contribute towards the regularization of such negative mass core thus making the interior geometry entirely regular. A detailed analysis of stellar geometries using an order-reduced RSET will appear in a forthcoming publication. 

Moving on to the regime where $\xi_{\rm c}<\xi<83/300$, we obtain solutions with a similar behaviour to that where $\xi\leq \xi_{\rm c}$ for the $h$ function as in Fig. \ref{Fig:Metrics}C, but the redshift $f$ now has a maximum at some $r>0$ and reaches a minimum at $r=0$. This is because the contributions within brackets in \eqref{Eq:fFieldEq} change sign for $\xi>\xi_{\rm c}$.

For $\xi>83/300$ we have a change in the behaviour of $h$. In this case, Eq.~\eqref{Eq:hFieldEq} now ensures that there will be no turning points in $h$ for any $r$. The function $f$ encounters again a turning point as the term within brackets in Eq.~\eqref{Eq:hFieldEq} is positive. Assuming $h$ vanishes at some radius $r_{0}$ as
\begin{equation}
    h=h_{1}(r-r_{0})+\order{r-r_{0}}^{2},
\end{equation}
and replacing this equation in Eq.~\eqref{Eq:hFieldEq} yields
\begin{equation}\label{Eq:hExpCase4}
    h_{1}=\frac{5l_{\rm P}^{2}}{r_{0}^{3}}\left(\xi-\frac{83}{300}\right).
\end{equation}
With the expansion \eqref{Eq:hExpCase4}, Eq. \eqref{Eq:fFieldEq} can be approximated by
\begin{equation}\label{Eq:RedsCase4}
    f'\simeq\chi_{1}f\left(r-r_{0}\right)^{-1},
\end{equation}
with 
\begin{equation}
    \chi_{1}=\frac{3\left(\xi-5/36\right)}{5\left(\xi-83/300\right)}.
\end{equation}
Solving Eq.~\eqref{Eq:RedsCase4} yields
\begin{equation}
    f\simeq\left(r-r_{0}\right)^{\chi_{1}}.
\end{equation}
Finally, the curvature scalar \eqref{Eq:Rscalar} diverges at $r_{0}$ as
\begin{equation}
    R\simeq\frac{\left(3-\chi_{1}\right)\chi_{1}}{2h_{1}\left(r-r_{0}\right)^{3}}.
\end{equation}

Figure \ref{Fig:Metrics}D shows a particular example of a solution belonging to this branch. For $\xi>83/300$, the resulting geometries display a null singularity. Since the ADM mass of the spacetime is negative, this singularity does not have the same physical relevance as the one that appeared in the positive ADM mass case from Fig. \ref{Fig:Metrics}A.

\section{Accuracy of the OR-RSET}\label{Sec:Validity}
An important aspect at this stage of the analysis is to what extent the solutions to the order-reduced system are good approximations. We undertake this analysis in what follows.

First, we need to establish criteria that define the ``proximity" between solutions to the order-reduced system
\begin{equation}\label{Eq:OrderRedEqs}
    G^{\mu}_{\nu}=8\pi\hbar\langle\hat{T}^{\mu}_{\nu}\rangle_{\rm{OR}}
\end{equation}
and solutions to the higher-derivative system \eqref{Eq:SemiEinstein}. We do so by defining
\begin{equation}\label{Eq:ValidityCrit}
    \mathcal{H}^{\mu}_{\nu}=G^{\mu}_{\nu}-8\pi\hbar\langle\hat{T}^{\mu}_{\nu}\rangle_{\text{AHS}},
\end{equation}
where $G^{\mu}_{\nu}$ and $\langle\hat{T}^{\mu}_{\nu}\rangle_{\text{AHS}}$ are calculated replacing the solutions $f$,~$h$ from~\eqref{Eq:OrderRedEqs},
and measuring the size of $\mathcal{H}^{\mu}_{\nu}$. As long as $\mathcal{H}^{\mu}_{\nu}=\order{\hbar^{2}}$, the terms $\order{\hbar^{2}}$ neglected in the expansions~\eqref{Eq:hbarExp} will correspond to subdominant contributions. Therefore, the solutions to Eq.~\eqref{Eq:OrderRedEqs} will amount to perturbative corrections of the solutions to Eq.~\eqref{Eq:ValidityCrit}, the discrepancy being given by the magnitude of the truncated terms.

\begin{figure*}
    \centering
    \includegraphics[width=\textwidth]{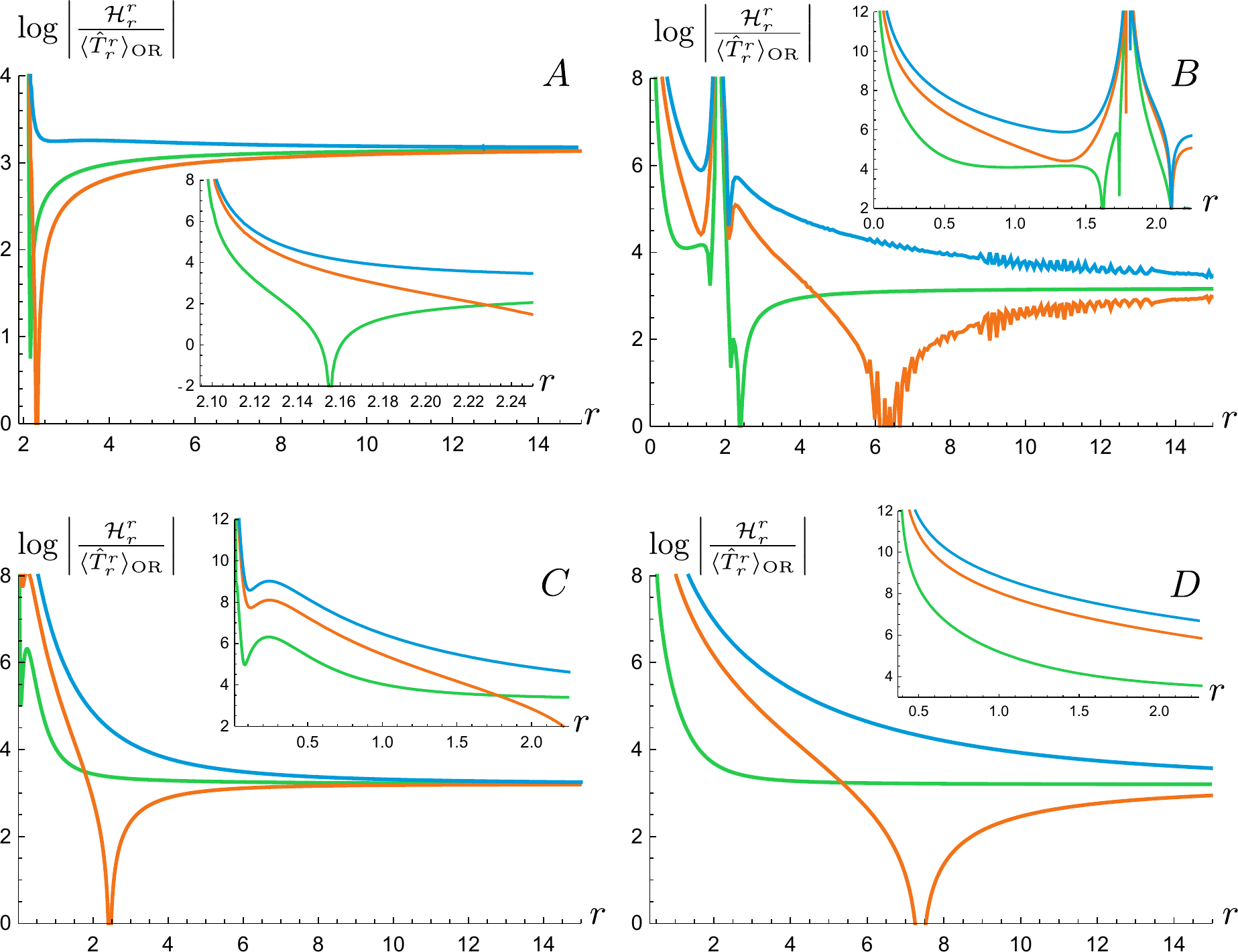}
    \caption{Plots of $\log{|\mathcal{H}^{r}_{r}/\langle\hat{T}^{r}_{r}\rangle_{\rm OR}|}$, a quantity measuring the deviation of the reduced order solutions with respect to the exact solutions to the higher-order semiclassical equations. 
    Panels {\it{A}} and {\it{B}} correspond to solutions with $M=1$ and $\xi=\{0,1\}$, respectively, whereas panels {\it{C}} and {\it{D}} correspond to solutions with $M=-1$ and $\xi=\{0,1\}$, respectively. As the AHS-RSET displays a free parameter $\nu$, we set $\nu=\{1,10^{-5},10^{10}\}$ for the green, orange and blue curves (the small oscillatory behaviour in panel {\it{B}} corresponds to numerical noise). We see that the validity of the solutions is not drastically affected by the value of $\nu$. The validity of these solutions gets progressively worse as $r$ diminishes, but the mismatch between solutions tends to the constant value $c\simeq3.17$ in the region of perturbative semiclassical corrections. }
    \label{Fig:Constraint}
\end{figure*}

We have examined the validity of solutions belonging to the regimes described in Section~\ref{Sec:Solutions} by plotting, in Fig.~\ref{Fig:Constraint}, the relative magnitude $\log|\mathcal{H}^{r}_{r}/\langle\hat{T}^{r}_{r}\rangle_{\rm OR}|$. The choice of the $rr$ component mitigates the error in numerically differentiating the solutions $f$ and $h$, since the highest-order terms in this quantity are proportional to $f^{(3)}$. Results show that the relative difference between the order-reduced solutions and the higher-order ones approaches a constant value at large distances, indicating that $\mathcal{H}^{r}_{r}$ decreases at the same rate as the component $\langle\hat{T}^{r}_{r}\rangle_{\rm OR}$, with a relative proportionality constant. The value of this constant is independent of the arbitrary renormalization scale $\nu$ appearing in the expressions from Appendix~\ref{Appendix:AHS}, but the quotient initially increases inwards for $\nu<1$ and decreases inwards for $\nu>1$. This is related to the change in sign of the $\log\nu$ terms appearing in the AHS-RSET. The quotient diverges at spacetime singularities or $r=0$, and independently of whether the spacetime is singular or regular. In the case of positive ADM mass geometries with $\xi\leq\xi_{\rm c}$, the quotient approaches a divergence at $r_{\rm D}$, precisely where the OR-RSET has dominant non-perturbative contributions. This is consistent with what we would expect from an order-reduced prescription.

\section{Comparison between approximations}
\label{Sec:Comparison}

The present work belongs to a broader investigation revolving around semiclassical backreaction effects in spherical symmetry. 
As a proxy to the study of more complicated field contents, it is customary to consider the propagation of a single massless and minimally coupled scalar field. Throughout the years, different approaches to modelling the RSET of minimal scalar fields have been developed~\cite{Polyakov1981,Andersonetal1995,FabbriNavarro-Salas2005,Arrecheaetal2020}. Given the difficulties in handling a potentially exact RSET, most of the approaches involve making approximations. All standard approximations properly capture effects associated with vacuum states at an event horizon (e.g. singularities at the event horizon in the zero-temperature state and thermal fluxes in nonzero-temperature ones). Preference of one over another depends on the specifics of the problem under consideration. 

However, despite their resemblance in what refers to non-local contributions, these approximations differ in their way of estimating purely local, curvature-dependent contributions. For example, the Polyakov approximation, coming from a dimensional reduction, lacks information about the behaviour of field modes at $r=0$. This missing information has to be provided by fixing the form of a free radial function [see Eq. \eqref{Eq:PolyakovRSET} below]. The AHS-RSET and the OR-RSET, on the other hand, are regular at $r=0$ (in the first case, under certain parity conditions for the metric, see Appendix \ref{Appendix:AHSReg}), and from this viewpoint can be considered an improvement with respect to Polyakov's approximation. However, as opposed to the Polyakov approximation, the AHS-RSET exhibits the free parameter $\nu$ (which accompanies purely local contributions) that affects the sign and magnitude of the RSET components.

Another discrepancy between these approximations (perhaps more crucial than the ambiguity in estimating local contributions) is that, given a fixed spacetime geometry, they can give place locally to more than one value for the RSET, even beyond the ambiguity in renormalization parameters.
As it is the case, then, the semiclassical equations can have branches of solutions that are not perturbatively connected to the classical solutions of the theory (the RSET does not go to zero in the $\hbar \to 0$ limit). These are what we call additional branches of solutions.
The introduction of additional branches is rooted to the differential structure of the RSET. To exemplify this point, consider the Regularized Polyakov RSET in~\cite{Arrecheaetal2022}
\begin{align}\label{Eq:PolyakovRSET}
    \langle\hat{T}^{t}_{t}\rangle=
    &
    F\frac{l_{\rm P}^{2}}{24\pi h}\left[\frac{2f'h'}{fh}+3\left(\frac{f'}{f}\right)^{2}-\frac{4f''}{f}\right],\nonumber\\
    \langle\hat{T}^{r}_{r}\rangle=
    &
    -F\frac{l_{\rm P}^{2}}{24\pi h}\left(\frac{f'}{f}\right)^{2},\nonumber\\
    \langle\hat{T}^{\theta}_{\theta}\rangle=
    &
    -\frac{l_{\rm P}^{2}}{48\pi h}\left(2F+rF'\right)\left(\frac{f'}{f}\right)^{2},
\end{align}
where $F$ is an arbitrary function of the radius that has to be fixed. These expressions exhibit a higher differential order than the OR-RSET~\eqref{Eq:VOR} in the temporal and radial components. 

Now, the $rr$ semiclassical equation \eqref{Eq:SemiEinstein} sourced by the Polyakov RSET~\eqref{Eq:PolyakovRSET} is
\begin{equation}\label{Eq:Polyakovrr}
    \frac{f+rf'-fh}{fhr^{2}}=-F\frac{l_{\rm P}^{2}}{3h}\left(\frac{f'}{f}\right)^{2},
\end{equation}
which can be recast into
\begin{equation}\label{Eq:Branches}
    f'=-\frac{3f}{2l_{\rm P}^{2}rF}\left[1\pm\sqrt{1+4l_{\rm P}^{2}F\left(h-1\right)/3}\right].
\end{equation}
Here, the branch with a negative sign (or unconcealed branch following the terminology in \cite{Arrecheaetal2020})  has a well-defined $l_{\rm P}\to0$ limit, therefore encoding both perturbative and non-perturbative corrections to the classical solutions. The branch with the positive sign (or concealed branch) does not have a well-defined classical limit and could be responsible for non-perturbative quantum effects. 
An important point is that these branches are not totally disconnected from each other.
We showed in \cite{Arrecheaetal2021} that jumps between branches take place in vacuum solutions {\it à la} Polyakov (although we did not find such jumps when dealing with stellar solutions~\cite{Arrecheaetal2022}); an integration that begins in the unconcealed branch can connect with the concealed one under certain conditions. 

The clearest example of this situation is the semiclassical Schwarzschild counterpart found in~\cite{Arrecheaetal2020} using the Regularized Polyakov approximation with \mbox{$F=1/(r^{2}+\alpha l_{\rm P}^{2})$} and $\alpha>1$. This is an asymmetric wormhole whose neck connects an asymptotically flat region with a new (singular) asymptotic region (note that to generate wormhole solutions with large necks one does not even need to regularize the Polyakov RSET~\cite{Fabbrietal2006}).
The asymptotically flat side of the wormhole lives within the unconcealed branch, whereas the singular side lives in the concealed one. Both branches are connected at the neck. The curvature singularity is null and located at finite affine distance from the neck (see \cite{Arrecheaetal2020} for a detailed analysis). In the more elaborate $s $-wave approximation that incorporates backscattering effects, the solution is also an asymmetric wormhole but the singularity is now timelike~\cite{Fabbrietal2006} (although appears to be at infinite radial distance as well~\cite{HoMatsuo2017}), which enhances the pathological nature of eternal vacuum solutions in semiclassical gravity. Recently, a similar analysis was done under an RSET approximation for conformally invariant fields~\cite{Beltran-Palau2022}, showing  remarkable agreement with the results obtained through Polyakov's approximation. In this case, the Schwarzschild counterpart also turns out to be an asymmetric wormhole displaying a null curvature singularity.

The AHS-RSET also contains plenty of non-perturbative branches of solutions. The only example we know of a non-perturbative solution to the semiclassical equations sourced by the AHS-RSET is~\cite{Hochbergetal1997}, where symmetric wormholes are obtained. These wormholes are not asymptotically flat, hence they cannot be compared with the asymptotically flat spacetimes that we obtained with the Polyakov approximation. 
On the contrary, the OR-RSET~\eqref{Eq:VOR} cannot introduce non-perturbative branches of solutions as they have been eliminated in the order reduction.

For instance, in the situation explored in the present work, we did not find wormhole solutions of any kind in the $\xi=0$ case. The geometry instead reaches a timelike (naked) singularity just above where the Schwarzschild radius would have been located (see Figs~\ref{Fig:Metrics}A and \ref{Fig:Deltar}). The same can be shown to happen by applying an order reduction to the Polyakov approximation. The resulting vacuum solution would result in a naked curvature singularity without the appearance of a wormhole neck.
What these analyses reveal is that the characteristics of self-consistent solutions in regions where non-perturbative effects kick in depends strongly on the physical content of the RSET approximation under consideration and its derivative order.
Nonetheless, different approximations to the RSET all lead to static vacuum solutions with positive ADM mass in which the event horizon has been substituted by a singularity of one type or another.

It is interesting to note here that preliminary explorations suggest that the situation is different when considering the more physical non-vacuum solutions. In~\cite{Arrecheaetal2022} we found semiclassical stellar configurations surpassing the Buchdahl compactness limit~\cite{Buchdahl1959}. These solutions are all found using just the perturbative (or unconcealed) branch. Preliminary analyses tell us that in this case very similar solutions are found using the OR-AHS. Confirmation of this idea would reinforce the robustness of said solutions.   

Our philosophy in here is not to argue for a particular approximation scheme as the best one; it is more to put all the possibilities on the table to see what they can offer. By adopting the view that the only trustable semiclassical self-consistent solutions are those perturbatively connected to a classical solution, the resulting solutions could turn out not to be very interesting, as these could not display any new qualitative behaviours. 
Instead, in this article our point of view is more heuristic and closer to the phenomenological philosophy underneath modified theories of gravity: motivating a possible form for some modifications of General Relativity and then analysing the new equations without caring how these equations might show up hierarchically from an even deeper description of spacetime.

\section{Conclusions and further discussion} 
\label{Sec:Conclusions}

In this work we have addressed the problem of semiclassical backreaction by reducing the order of the semiclassical equations in spherically symmetric vacuum situations. If said equations were tackled in full glory, solving them self-consistently would prove to be an extremely time consuming task due to their large space of solutions. In addition, there is no definite method for disregarding solutions based solely on their physical consistency.

Inspired by previous works in the literature \cite{ParkerSimon1993}, we have developed a procedure for obtaining regular, order-reduced RSET approximations that satisfy all the properties expected from a suitable RSET. We used this procedure to find the solutions to the semiclassical equations in vacuum. The results here obtained are consistent with previous analyses that make use of the Polyakov \cite{Arrecheaetal2020} and $s$-wave~\cite{Fabbrietal2003} approximations for the minimally coupled and massless scalar field in the region where semiclassical corrections are perturbative and the case $\xi=0$. Instead, in the region in which the semiclassical corrections are non-perturbative these works agree in the fact that the event horizon gets replaced by a curvature singularity in the Polyakov, $s$-wave and order-reduced approximations. 

For completion, we have analysed what happens for other $\xi$ couplings. When the coupling is increased beyond its separatrix value $\xi_{\rm c}=11/60$ the spacetime becomes regular and horizonless. However, curvature invariants become Planckian inside the region where the classical event horizon would have been located. For negative values of the ADM mass, there are both regular and irregular spacetimes depending on $\xi$. The regular ones could resemble the innermost regions of semiclassical relativistic stars that display negative-mass interiors generated by the effects of quantum vacuum polarization.

Following the procedure outlined in Sec. \ref{Sec:Order}, it is possible to obtain RSET approximations in various situations. The most straightforward extensions of this work would imply adding a classical electromagnetic field and/or a cosmological constant. In doing so, the validity of the very same AHS-RSET (from which the OR-RSET is constructed) requires careful evaluation~\cite{Andersonetal1995}. Ideally, any application of the AHS-RSET (or its order-reduced counterpart) should come along with a detailed analysis of the complete RSET incorporating the numerical part $\langle\hat{T}^{\mu}_{\nu}\rangle_{\text{num}}$, but this is technically challenging and escapes the scope of the current work.

The reduction of order that we have followed here is maximal in the sense that it eliminates all additional branches of solutions. However, a window remains open for the construction of partially order-reduced RSETs that, while yielding second-order equations of motion, still retain terms proportional to $f',~h',~f''$ and $h''$, in a similar fashion to the Polyakov approximation~\eqref{Eq:PolyakovRSET}. Is it possible to find a way to reduce the order of the AHS-RSET while retaining physically-meaningful branches of solutions? If so, how would this affect singularities? Presumably, while the non-perturbative characteristics revealed in vacuum solutions would be modified significantly, solutions that involve regular fluid spheres are more robust as the additional branches are not necessarily explored.

We have paved the path for the forthcoming analysis of semiclassical solutions incorporating a classical perfect fluid. The authors have proved that semiclassical effects in the form of a modified Polyakov approximation allow for the existence of regular stars that surpass the Buchdahl compactness limit~\cite{Arrechea2021b}. Preliminary investigations indicate that approximations to the RSET based on a reduction of order also generate configurations with akin characteristics. The fact that different modelings of the semiclassical equations end up describing similar physical scenarios provides an extra degree of robustness to semiclassical analyses.


\section{Acknowledgements}
The authors thank Valentin Boyanov and Gerardo Garc\'{\i}a-Moreno for very useful discussions. Financial support was provided by the Spanish Government through the projects PID2020-118159GB-C43/AEI/10.13039/501100011033, PID2020-118159GB-C44/AEI/10.13039/501100011033, PID2019-107847RB-C44/AEI/10.13039/501100011033, and by the Junta de Andalucía through the project FQM219. This research is supported by a research grant (29405) from VILLUM fonden. CB and JA acknowledges financial support from the State Agency for Research of the Spanish MCIU through the “Center of Excellence Severo Ochoa” award to the Instituto de Astrofísica de Andalucía (SEV-2017-0709).

\appendix

\section{The Anderson-Hiscock-Samuel RSET}
\label{Appendix:AHS}
Below we show the components of the AHS-RSET for a massless field with arbitrary coupling $\xi$ (temperature-dependent terms are retained). Comparing these expressions with those of the OR-RSET, we observe that the simplification is dramatic.
\begin{widetext}
\begin{dmath*}
1440\pi^{2}\langle\hat{T}^{t}_{t}\rangle_{\rm AHS}=
-\frac{3 \kappa ^4}{f^2}+\left(\frac{75 \left(f'\right)^2}{f^3 h}+\frac{30 h' f'}{f^2 h^2}-\frac{120 f'}{f^2 h r}+\frac{30 h'}{f h^2 r}-\frac{60 f''}{f^2 h}+\frac{30}{f r^2}-\frac{30}{f h r^2}\right) \left(\xi -\frac{1}{6}\right) \kappa ^2+\frac{7 \left(f'\right)^4}{32 f^4 h^2}+\frac{7 \left(h'\right)^3}{h^5 r}+\frac{3 \left(f'\right)^2}{4 f^2 h^2 r^2}+\frac{5 f' h'}{2 f h^3 r^2}+\frac{\left(f'\right)^2 f''}{8 f^3 h^2}+\frac{19 \left(h'\right)^2 f''}{8 f h^4}+\frac{9 f' h' f''}{8 f^2 h^3}+\frac{\left(f'\right)^2 h''}{4 f^2 h^3}+\frac{f' h''}{2 f h^3 r}+\frac{13 f' h' h''}{8 f h^4}+\frac{h''}{h^3 r^2}+\frac{2 f^{(3)}}{f h^2 r}+\frac{h^{(3)}}{h^3 r}+\frac{f^{(4)}}{2 f h^2}-\frac{f' f^{(3)}}{2 f^2 h^2}-\frac{3 \left(f''\right)^2}{8 f^2 h^2}-\frac{f'' h''}{f h^3}-\frac{3 h' f^{(3)}}{2 f h^3}-\frac{f' h^{(3)}}{4 f h^3}-\frac{\left(f'\right)^3 h'}{16 f^3 h^3}-\frac{19 \left(f'\right)^2 \left(h'\right)^2}{32 f^2 h^4}-\frac{7 f' \left(h'\right)^3}{4 f h^5}-\frac{3 \left(f'\right)^3}{4 f^3 h^2 r}-\frac{2 h' f''}{f h^3 r}-\frac{5 \left(f'\right)^2 h'}{4 f^2 h^3 r}-\frac{13 h' h''}{2 h^4 r}-\frac{3 f' \left(h'\right)^2}{2 f h^4 r}-\frac{7 \left(h'\right)^2}{4 h^4 r^2}-\frac{2 h'}{h^3 r^3}+\frac{1}{r^4}-\frac{1}{h^2 r^4}+\left(-\frac{49 \left(f'\right)^4}{32 f^4 h^2}-\frac{29 h' \left(f'\right)^3}{16 f^3 h^3}+\frac{11 \left(f'\right)^3}{8 f^3 h^2 r}+\frac{3 h' \left(f'\right)^2}{2 f^2 h^3 r}+\frac{29 f'' \left(f'\right)^2}{8 f^3 h^2}+\frac{3 h'' \left(f'\right)^2}{4 f^2 h^3}-\frac{57 \left(h'\right)^2 \left(f'\right)^2}{32 f^2 h^4}+\frac{5 \left(f'\right)^2}{8 f^2 h^2 r^2}+\frac{13 \left(h'\right)^2 f'}{8 f h^4 r}+\frac{5 h' f'}{4 f h^3 r^2}+\frac{27 h' f'' f'}{8 f^2 h^3}+\frac{13 h' h'' f'}{8 f h^4}-\frac{3 f^{(3)} f'}{2 f^2 h^2}-\frac{h^{(3)} f'}{4 f h^3}-\frac{7 \left(h'\right)^3 f'}{4 f h^5}-\frac{13 f'' f'}{4 f^2 h^2 r}-\frac{3 h'' f'}{4 f h^3 r}+\frac{7 \left(h'\right)^3}{2 h^5 r}+\frac{19 \left(h'\right)^2 f''}{8 f h^4}+\frac{h''}{2 h^3 r^2}+\frac{2 f^{(3)}}{f h^2 r}+\frac{h^{(3)}}{2 h^3 r}+\frac{f^{(4)}}{2 f h^2}-\frac{9 \left(f''\right)^2}{8 f^2 h^2}-\frac{f'' h''}{f h^3}-\frac{3 h' f^{(3)}}{2 f h^3}-\frac{13 h' f''}{4 f h^3 r}-\frac{13 h' h''}{4 h^4 r}-\frac{7 \left(h'\right)^2}{8 h^4 r^2}-\frac{h'}{h^3 r^3}+\frac{1}{2 r^4}-\frac{1}{2 h^2 r^4}\right) \log f+\left(-\frac{49 \left(f'\right)^4}{16 f^4 h^2}-\frac{29 h' \left(f'\right)^3}{8 f^3 h^3}+\frac{11 \left(f'\right)^3}{4 f^3 h^2 r}+\frac{3 h' \left(f'\right)^2}{f^2 h^3 r}+\frac{29 f'' \left(f'\right)^2}{4 f^3 h^2}+\frac{3 h'' \left(f'\right)^2}{2 f^2 h^3}-\frac{57 \left(h'\right)^2 \left(f'\right)^2}{16 f^2 h^4}+\frac{5 \left(f'\right)^2}{4 f^2 h^2 r^2}+\frac{13 \left(h'\right)^2 f'}{4 f h^4 r}+\frac{5 h' f'}{2 f h^3 r^2}+\frac{27 h' f'' f'}{4 f^2 h^3}+\frac{13 h' h'' f'}{4 f h^4}-\frac{3 f^{(3)} f'}{f^2 h^2}-\frac{h^{(3)} f'}{2 f h^3}-\frac{7 \left(h'\right)^3 f'}{2 f h^5}-\frac{13 f'' f'}{2 f^2 h^2 r}-\frac{3 h'' f'}{2 f h^3 r}+\frac{7 \left(h'\right)^3}{h^5 r}+\frac{19 \left(h'\right)^2 f''}{4 f h^4}+\frac{h''}{h^3 r^2}+\frac{4 f^{(3)}}{f h^2 r}+\frac{h^{(3)}}{h^3 r}+\frac{f^{(4)}}{f h^2}-\frac{9 \left(f''\right)^2}{4 f^2 h^2}-\frac{2 f'' h''}{f h^3}-\frac{3 h' f^{(3)}}{f h^3}-\frac{13 h' f''}{2 f h^3 r}-\frac{13 h' h''}{2 h^4 r}-\frac{7 \left(h'\right)^2}{4 h^4 r^2}-\frac{2 h'}{h^3 r^3}+\frac{1}{r^4}-\frac{1}{h^2 r^4}\right) \log \nu\end{dmath*}
\begin{dmath}\label{Eq:AHSRSETtt}
+\left(-\frac{945 \left(f'\right)^4}{8 f^4 h^2}-\frac{495 h' \left(f'\right)^3}{4 f^3 h^3}+\frac{165 \left(f'\right)^3}{f^3 h^2 r}+\frac{405 h' \left(f'\right)^2}{2 f^2 h^3 r}+\frac{495 f'' \left(f'\right)^2}{2 f^3 h^2}+\frac{45 h'' \left(f'\right)^2}{f^2 h^3}-\frac{855 \left(h'\right)^2 \left(f'\right)^2}{8 f^2 h^4}+\frac{45 \left(f'\right)^2}{2 f^2 h r^2}-\frac{45 \left(f'\right)^2}{2 f^2 h^2 r^2}+\frac{285 \left(h'\right)^2 f'}{f h^4 r}+\frac{15 h' f'}{f h^2 r^2}+\frac{405 h' f'' f'}{2 f^2 h^3}+\frac{195 h' h'' f'}{2 f h^4}-\frac{90 f^{(3)} f'}{f^2 h^2}-\frac{15 h^{(3)} f'}{f h^3}-\frac{105 \left(h'\right)^3 f'}{f h^5}-\frac{270 f'' f'}{f^2 h^2 r}-\frac{120 h'' f'}{f h^3 r}-\frac{45 h' f'}{f h^3 r^2}+\frac{60 \left(h'\right)^2}{h^4 r^2}+\frac{90 h'}{h^3 r^3}+\frac{285 \left(h'\right)^2 f''}{2 f h^4}+\frac{30 f''}{f h^2 r^2}+\frac{195 h' h''}{h^4 r}+\frac{120 f^{(3)}}{f h^2 r}+\frac{30 f^{(4)}}{f h^2}-\frac{135 \left(f''\right)^2}{2 f^2 h^2}-\frac{60 f'' h''}{f h^3}-\frac{90 h' f^{(3)}}{f h^3}-\frac{30 h^{(3)}}{h^3 r}-\frac{270 h' f''}{f h^3 r}-\frac{210 \left(h'\right)^3}{h^5 r}-\frac{30 f''}{f h r^2}-\frac{30 h''}{h^3 r^2}-\frac{30 h'}{h^2 r^3}-\frac{60}{h r^4}+\frac{60}{h^2 r^4}\right) \left(\xi -\frac{1}{6}\right)
+\left[\frac{945 \left(f'\right)^4}{4 f^4 h^2}+\frac{270 h' \left(f'\right)^3}{f^3 h^3}-\frac{720 \left(f'\right)^3}{f^3 h^2 r}+\frac{855 \left(h'\right)^2 \left(f'\right)^2}{4 f^2 h^4}-\frac{540 f'' \left(f'\right)^2}{f^3 h^2}-\frac{90 h'' \left(f'\right)^2}{f^2 h^3}-\frac{360 h' \left(f'\right)^2}{f^2 h^3 r}+\frac{270 \left(f'\right)^2}{f^2 h r^2}-\frac{90 \left(f'\right)^2}{f^2 h^2 r^2}+\frac{900 \left(h'\right)^2 f'}{f h^4 r}+\frac{180 h' f'}{f h^2 r^2}+\frac{900 f'' f'}{f^2 h^2 r}+\frac{180 f^{(3)} f'}{f^2 h^2}-\frac{405 h' f'' f'}{f^2 h^3}-\frac{360 h'' f'}{f h^3 r}-\frac{900 h' f'}{f h^3 r^2}+\frac{180 \left(h'\right)^2}{h^4 r^2}+\frac{135 \left(f''\right)^2}{f^2 h^2}+\frac{360 h'}{h^2 r^3}+\frac{360 f''}{f h^2 r^2}-\frac{360 h' f''}{f h^3 r}-\frac{360 f''}{f h r^2}-\frac{360 h'}{h^3 r^3}+\frac{180}{r^4}-\frac{360}{h r^4}+\frac{180}{h^2 r^4}+\left(-\frac{2205 \left(f'\right)^4}{8 f^4 h^2}-\frac{1305 h' \left(f'\right)^3}{4 f^3 h^3}+\frac{585 \left(f'\right)^3}{f^3 h^2 r}+\frac{675 h' \left(f'\right)^2}{f^2 h^3 r}+\frac{1305 f'' \left(f'\right)^2}{2 f^3 h^2}+\frac{135 h'' \left(f'\right)^2}{f^2 h^3}-\frac{2565 \left(h'\right)^2 \left(f'\right)^2}{8 f^2 h^4}-\frac{90 \left(f'\right)^2}{f^2 h^2 r^2}+\frac{630 \left(h'\right)^2 f'}{f h^4 r}+\frac{90 h' f'}{f h^3 r^2}+\frac{1215 h' f'' f'}{2 f^2 h^3}+\frac{585 h' h'' f'}{2 f h^4}-\frac{270 f^{(3)} f'}{f^2 h^2}-\frac{45 h^{(3)} f'}{f h^3}-\frac{315 \left(h'\right)^3 f'}{f h^5}-\frac{990 f'' f'}{f^2 h^2 r}-\frac{270 h'' f'}{f h^3 r}+\frac{450 \left(h'\right)^2}{h^4 r^2}+\frac{360 h'}{h^3 r^3}+\frac{855 \left(h'\right)^2 f''}{2 f h^4}+\frac{1170 h' h''}{h^4 r}+\frac{360 f^{(3)}}{f h^2 r}+\frac{90 f^{(4)}}{f h^2}-\frac{405 \left(f''\right)^2}{2 f^2 h^2}-\frac{180 f'' h''}{f h^3}-\frac{270 h' f^{(3)}}{f h^3}-\frac{180 h^{(3)}}{h^3 r}-\frac{720 h' f''}{f h^3 r}-\frac{1260 \left(h'\right)^3}{h^5 r}-\frac{180 h''}{h^3 r^2}+\frac{90}{r^4}-\frac{540}{h r^4}+\frac{450}{h^2 r^4}\right) \log f+\left(-\frac{2205 \left(f'\right)^4}{4 f^4 h^2}-\frac{1305 h' \left(f'\right)^3}{2 f^3 h^3}+\frac{1170 \left(f'\right)^3}{f^3 h^2 r}+\frac{1350 h' \left(f'\right)^2}{f^2 h^3 r}+\frac{1305 f'' \left(f'\right)^2}{f^3 h^2}+\frac{270 h'' \left(f'\right)^2}{f^2 h^3}-\frac{2565 \left(h'\right)^2 \left(f'\right)^2}{4 f^2 h^4}-\frac{180 \left(f'\right)^2}{f^2 h^2 r^2}+\frac{1260 \left(h'\right)^2 f'}{f h^4 r}+\frac{180 h' f'}{f h^3 r^2}+\frac{1215 h' f'' f'}{f^2 h^3}+\frac{585 h' h'' f'}{f h^4}-\frac{540 f^{(3)} f'}{f^2 h^2}-\frac{90 h^{(3)} f'}{f h^3}-\frac{630 \left(h'\right)^3 f'}{f h^5}-\frac{1980 f'' f'}{f^2 h^2 r}-\frac{540 h'' f'}{f h^3 r}+\frac{900 \left(h'\right)^2}{h^4 r^2}+\frac{720 h'}{h^3 r^3}+\frac{855 \left(h'\right)^2 f''}{f h^4}+\frac{2340 h' h''}{h^4 r}+\frac{720 f^{(3)}}{f h^2 r}+\frac{180 f^{(4)}}{f h^2}-\frac{405 \left(f''\right)^2}{f^2 h^2}-\frac{360 f'' h''}{f h^3}-\frac{540 h' f^{(3)}}{f h^3}-\frac{360 h^{(3)}}{h^3 r}-\frac{1440 h' f''}{f h^3 r}-\frac{2520 \left(h'\right)^3}{h^5 r}-\frac{360 h''}{h^3 r^2}+\frac{180}{r^4}-\frac{1080}{h r^4}+\frac{900}{h^2 r^4}\right) \log \nu\right] \left(\xi -\frac{1}{6}\right)^2,
\end{dmath}
\begin{dmath}\label{Eq:AHSRSETrr}
1440\pi^{2}\langle\hat{T}^{r}_{r}\rangle_{\rm AHS}=\frac{\kappa ^4}{f^2}+\left(-\frac{15 \left(f'\right)^2}{f^3 h}-\frac{30 f'}{f^2 h r}-\frac{30}{f r^2}+\frac{30}{f h r^2}\right) \left(\xi -\frac{1}{6}\right) \kappa ^2+\frac{\left(f'\right)^4}{32 f^4 h^2}+\frac{7 \left(f'\right)^2 \left(h'\right)^2}{32 f^2 h^4}+\frac{2 f'}{f h^2 r^3}+\frac{\left(f'\right)^3 h'}{16 f^3 h^3}+\frac{f' h'}{2 f h^3 r^2}+\frac{2 f' f''}{f^2 h^2 r}+\frac{h' f''}{f h^3 r}+\frac{f' h''}{f h^3 r}+\frac{f' f^{(3)}}{4 f^2 h^2}-\frac{\left(f''\right)^2}{8 f^2 h^2}-\frac{\left(f'\right)^2 f''}{8 f^3 h^2}-\frac{f' h' f''}{4 f^2 h^3}-\frac{\left(f'\right)^2 h''}{8 f^2 h^3}-\frac{f^{(3)}}{f h^2 r}-\frac{\left(f'\right)^3}{2 f^3 h^2 r}-\frac{\left(f'\right)^2 h'}{4 f^2 h^3 r}-\frac{7 f' \left(h'\right)^2}{4 f h^4 r}-\frac{2 f''}{f h^2 r^2}+\left(\frac{7 \left(f'\right)^4}{32 f^4 h^2}+\frac{3 h' \left(f'\right)^3}{16 f^3 h^3}-\frac{5 \left(f'\right)^3}{8 f^3 h^2 r}+\frac{7 \left(h'\right)^2 \left(f'\right)^2}{32 f^2 h^4}-\frac{3 f'' \left(f'\right)^2}{8 f^3 h^2}-\frac{h'' \left(f'\right)^2}{8 f^2 h^3}-\frac{h' \left(f'\right)^2}{2 f^2 h^3 r}-\frac{\left(f'\right)^2}{8 f^2 h^2 r^2}+\frac{h' f'}{4 f h^3 r^2}+\frac{3 f'' f'}{2 f^2 h^2 r}+\frac{h'' f'}{2 f h^3 r}+\frac{f^{(3)} f'}{4 f^2 h^2}-\frac{h' f'' f'}{4 f^2 h^3}-\frac{7 \left(h'\right)^2 f'}{8 f h^4 r}+\frac{f'}{f h^2 r^3}+\frac{7 \left(h'\right)^2}{8 h^4 r^2}+\frac{h' f''}{2 f h^3 r}-\frac{\left(f''\right)^2}{8 f^2 h^2}-\frac{f^{(3)}}{2 f h^2 r}-\frac{f''}{f h^2 r^2}-\frac{h''}{2 h^3 r^2}+\frac{1}{2 r^4}-\frac{1}{2 h^2 r^4}\right) \log f+\left(\frac{7 \left(f'\right)^4}{16 f^4 h^2}+\frac{3 h' \left(f'\right)^3}{8 f^3 h^3}-\frac{5 \left(f'\right)^3}{4 f^3 h^2 r}+\frac{7 \left(h'\right)^2 \left(f'\right)^2}{16 f^2 h^4}-\frac{3 f'' \left(f'\right)^2}{4 f^3 h^2}-\frac{h'' \left(f'\right)^2}{4 f^2 h^3}-\frac{h' \left(f'\right)^2}{f^2 h^3 r}-\frac{\left(f'\right)^2}{4 f^2 h^2 r^2}+\frac{h' f'}{2 f h^3 r^2}+\frac{3 f'' f'}{f^2 h^2 r}+\frac{h'' f'}{f h^3 r}+\frac{f^{(3)} f'}{2 f^2 h^2}-\frac{h' f'' f'}{2 f^2 h^3}-\frac{7 \left(h'\right)^2 f'}{4 f h^4 r}+\frac{2 f'}{f h^2 r^3}+\frac{7 \left(h'\right)^2}{4 h^4 r^2}+\frac{h' f''}{f h^3 r}-\frac{\left(f''\right)^2}{4 f^2 h^2}-\frac{f^{(3)}}{f h^2 r}-\frac{2 f''}{f h^2 r^2}-\frac{h''}{h^3 r^2}+\frac{1}{r^4}-\frac{1}{h^2 r^4}\right) \log\nu
+\left(\frac{135 \left(f'\right)^4}{8 f^4 h^2}+\frac{45 h' \left(f'\right)^3}{4 f^3 h^3}+\frac{45 \left(f'\right)^3}{f^3 h^2 r}+\frac{105 \left(h'\right)^2 \left(f'\right)^2}{8 f^2 h^4}-\frac{45 f'' \left(f'\right)^2}{2 f^3 h^2}-\frac{15 h'' \left(f'\right)^2}{2 f^2 h^3}-\frac{15 \left(f'\right)^2}{2 f^2 h r^2}-\frac{165 \left(f'\right)^2}{2 f^2 h^2 r^2}+\frac{105 \left(h'\right)^2 f'}{2 f h^4 r}+\frac{15 f^{(3)} f'}{f^2 h^2}-\frac{15 h' f'' f'}{f^2 h^3}-\frac{75 f'' f'}{f^2 h^2 r}-\frac{30 h'' f'}{f h^3 r}-\frac{30 h' f'}{f h^3 r^2}+\frac{30 f'}{f h r^3}-\frac{90 f'}{f h^2 r^3}+\frac{60 f''}{f h^2 r^2}+\frac{30 f^{(3)}}{f h^2 r}-\frac{15 \left(f''\right)^2}{2 f^2 h^2}-\frac{30 h' f''}{f h^3 r}\right) \left(\xi -\frac{1}{6}\right)+\left[-\frac{45 \left(f'\right)^4}{2 f^4 h^2}-\frac{45 h' \left(f'\right)^3}{2 f^3 h^3}+\frac{45 f'' \left(f'\right)^2}{f^3 h^2}-\frac{180 h' \left(f'\right)^2}{f^2 h^3 r}-\frac{90 \left(f'\right)^2}{f^2 h r^2}+\frac{450 \left(f'\right)^2}{f^2 h^2 r^2}+\frac{180 f'' f'}{f^2 h^2 r}-\frac{360 h' f'}{f h^3 r^2}-\frac{360 f'}{f h r^3}+\frac{360 f'}{f h^2 r^3}+\left(\frac{315 \left(f'\right)^4}{8 f^4 h^2}+\frac{135 h' \left(f'\right)^3}{4 f^3 h^3}+\frac{90 \left(f'\right)^3}{f^3 h^2 r}+\frac{315 \left(h'\right)^2 \left(f'\right)^2}{8 f^2 h^4}+\frac{45 h' \left(f'\right)^2}{f^2 h^3 r}-\frac{135 f'' \left(f'\right)^2}{2 f^3 h^2}-\frac{45 h'' \left(f'\right)^2}{2 f^2 h^3}-\frac{360 \left(f'\right)^2}{f^2 h^2 r^2}+\frac{315 \left(h'\right)^2 f'}{f h^4 r}+\frac{45 f^{(3)} f'}{f^2 h^2}-\frac{45 h' f'' f'}{f^2 h^3}-\frac{270 f'' f'}{f^2 h^2 r}-\frac{180 h'' f'}{f h^3 r}-\frac{360 h' f'}{f h^3 r^2}-\frac{360 f'}{f h^2 r^3}+\frac{630 \left(h'\right)^2}{h^4 r^2}+\frac{360 f''}{f h^2 r^2}+\frac{180 f^{(3)}}{f h^2 r}-\frac{45 \left(f''\right)^2}{2 f^2 h^2}-\frac{180 h' f''}{f h^3 r}-\frac{360 h''}{h^3 r^2}+\frac{90}{r^4}+\frac{540}{h r^4}-\frac{630}{h^2 r^4}\right) \log f+\left(\frac{315 \left(f'\right)^4}{4 f^4 h^2}+\frac{135 h' \left(f'\right)^3}{2 f^3 h^3}+\frac{180 \left(f'\right)^3}{f^3 h^2 r}+\frac{315 \left(h'\right)^2 \left(f'\right)^2}{4 f^2 h^4}+\frac{90 h' \left(f'\right)^2}{f^2 h^3 r}-\frac{135 f'' \left(f'\right)^2}{f^3 h^2}-\frac{45 h'' \left(f'\right)^2}{f^2 h^3}-\frac{720 \left(f'\right)^2}{f^2 h^2 r^2}+\frac{630 \left(h'\right)^2 f'}{f h^4 r}+\frac{90 f^{(3)} f'}{f^2 h^2}-\frac{90 h' f'' f'}{f^2 h^3}-\frac{540 f'' f'}{f^2 h^2 r}-\frac{360 h'' f'}{f h^3 r}-\frac{720 h' f'}{f h^3 r^2}-\frac{720 f'}{f h^2 r^3}+\frac{1260 \left(h'\right)^2}{h^4 r^2}\\
+\frac{720 f''}{f h^2 r^2}+\frac{360 f^{(3)}}{f h^2 r}-\frac{45 \left(f''\right)^2}{f^2 h^2}-\frac{360 h' f''}{f h^3 r}-\frac{720 h''}{h^3 r^2}+\frac{180}{r^4}+\frac{1080}{h r^4}-\frac{1260}{h^2 r^4}\right) \log\nu\right] \left(\xi -\frac{1}{6}\right)^2,
\end{dmath}
\begin{dmath*}
1440\pi^{2}\langle\hat{T}^{\theta}_{\theta}\rangle_{\rm AHS}=\frac{\kappa ^4}{f^2}+\left(\frac{75 \left(f'\right)^2}{2 f^3 h}+\frac{15 h' f'}{2 f^2 h^2}-\frac{15 f'}{f^2 h r}-\frac{15 f''}{f^2 h}-\frac{15 h'}{f h^2 r}\right) \left(\xi -\frac{1}{6}\right) \kappa ^2+\frac{17 \left(f'\right)^4}{32 f^4 h^2}+\frac{7 f' \left(h'\right)^3}{4 f h^5}+\frac{31 \left(f'\right)^2 \left(h'\right)^2}{32 f^2 h^4}+\frac{7 \left(f''\right)^2}{8 f^2 h^2}+\frac{13 \left(f'\right)^3 h'}{16 f^3 h^3}+\frac{f' f''}{f^2 h^2 r}+\frac{11 h' f''}{4 f h^3 r}+\frac{f''}{f h^2 r^2}+\frac{f' h''}{4 f h^3 r}+\frac{f'' h''}{f h^3}+\frac{3 f' f^{(3)}}{4 f^2 h^2}+\frac{3 h' f^{(3)}}{2 f h^3}+\frac{f' h^{(3)}}{4 f h^3}-\frac{f^{(4)}}{2 f h^2}-\frac{13 \left(f'\right)^2 f''}{8 f^3 h^2}-\frac{2 f' h' f''}{f^2 h^3}-\frac{3 \left(f'\right)^2 h''}{8 f^2 h^3}-\frac{19 \left(h'\right)^2 f''}{8 f h^4}-\frac{13 f' h' h''}{8 f h^4}-\frac{3 f^{(3)}}{2 f h^2 r}-\frac{\left(f'\right)^3}{2 f^3 h^2 r}-\frac{3 \left(f'\right)^2 h'}{4 f^2 h^3 r}-\frac{3 f' \left(h'\right)^2}{4 f h^4 r}-\frac{3 f' h'}{2 f h^3 r^2}-\frac{f'}{f h^2 r^3}+\left(\frac{21 \left(f'\right)^4}{32 f^4 h^2}+\frac{13 h' \left(f'\right)^3}{16 f^3 h^3}-\frac{3 \left(f'\right)^3}{8 f^3 h^2 r}+\frac{25 \left(h'\right)^2 \left(f'\right)^2}{32 f^2 h^4}-\frac{13 f'' \left(f'\right)^2}{8 f^3 h^2}-\frac{5 h'' \left(f'\right)^2}{16 f^2 h^3}-\frac{h' \left(f'\right)^2}{2 f^2 h^3 r}-\frac{\left(f'\right)^2}{4 f^2 h^2 r^2}+\frac{7 \left(h'\right)^3 f'}{8 f h^5}+\frac{7 f'' f'}{8 f^2 h^2 r}+\frac{h'' f'}{8 f h^3 r}+\frac{5 f^{(3)} f'}{8 f^2 h^2}+\frac{h^{(3)} f'}{8 f h^3}-\frac{25 h' f'' f'}{16 f^2 h^3}-\frac{13 h' h'' f'}{16 f h^4}-\frac{3 \left(h'\right)^2 f'}{8 f h^4 r}-\frac{3 h' f'}{4 f h^3 r^2}-\frac{f'}{2 f h^2 r^3}+\frac{5 \left(f''\right)^2}{8 f^2 h^2}+\frac{h'}{2 h^3 r^3}+\frac{11 h' f''}{8 f h^3 r}+\frac{f''}{2 f h^2 r^2}+\frac{13 h' h''}{8 h^4 r}+\frac{f'' h''}{2 f h^3}+\frac{3 h' f^{(3)}}{4 f h^3}-\frac{f^{(4)}}{4 f h^2}-\frac{19 \left(h'\right)^2 f''}{16 f h^4}-\frac{3 f^{(3)}}{4 f h^2 r}-\frac{h^{(3)}}{4 h^3 r}-\frac{7 \left(h'\right)^3}{4 h^5 r}-\frac{1}{2 r^4}+\frac{1}{2 h^2 r^4}\right) \log f+\left(\frac{21 \left(f'\right)^4}{16 f^4 h^2}+\frac{13 h' \left(f'\right)^3}{8 f^3 h^3}-\frac{3 \left(f'\right)^3}{4 f^3 h^2 r}+\frac{25 \left(h'\right)^2 \left(f'\right)^2}{16 f^2 h^4}-\frac{13 f'' \left(f'\right)^2}{4 f^3 h^2}-\frac{5 h'' \left(f'\right)^2}{8 f^2 h^3}-\frac{h' \left(f'\right)^2}{f^2 h^3 r}-\frac{\left(f'\right)^2}{2 f^2 h^2 r^2}+\frac{7 \left(h'\right)^3 f'}{4 f h^5}+\frac{7 f'' f'}{4 f^2 h^2 r}+\frac{h'' f'}{4 f h^3 r}+\frac{5 f^{(3)} f'}{4 f^2 h^2}+\frac{h^{(3)} f'}{4 f h^3}-\frac{25 h' f'' f'}{8 f^2 h^3}-\frac{13 h' h'' f'}{8 f h^4}-\frac{3 \left(h'\right)^2 f'}{4 f h^4 r}-\frac{3 h' f'}{2 f h^3 r^2}-\frac{f'}{f h^2 r^3}+\frac{5 \left(f''\right)^2}{4 f^2 h^2}+\frac{h'}{h^3 r^3}+\frac{11 h' f''}{4 f h^3 r}+\frac{f''}{f h^2 r^2}+\frac{13 h' h''}{4 h^4 r}+\frac{f'' h''}{f h^3}+\frac{3 h' f^{(3)}}{2 f h^3}-\frac{f^{(4)}}{2 f h^2}-\frac{19 \left(h'\right)^2 f''}{8 f h^4}-\frac{3 f^{(3)}}{2 f h^2 r}-\frac{h^{(3)}}{2 h^3 r}-\frac{7 \left(h'\right)^3}{2 h^5 r}-\frac{1}{r^4}+\frac{1}{h^2 r^4}\right) \log\nu+\left(-\frac{645 \left(f'\right)^4}{8 f^4 h^2}-\frac{675 h' \left(f'\right)^3}{8 f^3 h^3}+\frac{90 \left(f'\right)^3}{f^3 h^2 r}+\frac{405 h' \left(f'\right)^2}{4 f^2 h^3 r}+\frac{675 f'' \left(f'\right)^2}{4 f^3 h^2}+\frac{30 h'' \left(f'\right)^2}{f^2 h^3}-\frac{285 \left(h'\right)^2 \left(f'\right)^2}{4 f^2 h^4}-\frac{15 \left(f'\right)^2}{2 f^2 h r^2}+\frac{45 \left(f'\right)^2}{2 f^2 h^2 r^2}+\frac{225 \left(h'\right)^2 f'}{4 f h^4 r}+\frac{135 h' f'}{2 f h^3 r^2}+\frac{135 h' f'' f'}{f^2 h^3}+\frac{195 h' h'' f'}{4 f h^4}-\frac{60 f^{(3)} f'}{f^2 h^2}-\frac{15 h^{(3)} f'}{2 f h^3}-\frac{105 \left(h'\right)^3 f'}{2 f h^5}-\frac{285 f'' f'}{2 f^2 h^2 r}-\frac{45 h'' f'}{2 f h^3 r}-\frac{15 h' f'}{2 f h^2 r^2}+\frac{30 f'}{f h^2 r^3}+\frac{285 \left(h'\right)^2 f''}{4 f h^4}+\frac{15 f''}{f h r^2}+\frac{45 f^{(3)}}{f h^2 r}+\frac{15 f^{(4)}}{f h^2}-\frac{45 \left(f''\right)^2}{f^2 h^2}-\frac{30 f'' h''}{f h^3}-\frac{45 h' f^{(3)}}{f h^3}-\frac{90 h' f''}{f h^3 r}-\frac{45 f''}{f h^2 r^2}\right) \left(\xi -\frac{1}{6}\right)
\end{dmath*}
\begin{dmath}\label{Eq:AHSRSETthth}
+\left[\frac{405 \left(f'\right)^4}{2 f^4 h^2}+\frac{225 h' \left(f'\right)^3}{f^3 h^3}-\frac{495 \left(f'\right)^3}{f^3 h^2 r}+\frac{405 \left(h'\right)^2 \left(f'\right)^2}{2 f^2 h^4}-\frac{450 f'' \left(f'\right)^2}{f^3 h^2}-\frac{90 h'' \left(f'\right)^2}{f^2 h^3}-\frac{405 h' \left(f'\right)^2}{f^2 h^3 r}+\frac{90 \left(f'\right)^2}{f^2 h r^2}-\frac{270 \left(f'\right)^2}{f^2 h^2 r^2}+\frac{810 \left(h'\right)^2 f'}{f h^4 r}+\frac{90 h' f'}{f h^2 r^2}+\frac{630 f'' f'}{f^2 h^2 r}+\frac{180 f^{(3)} f'}{f^2 h^2}-\frac{360 h' f'' f'}{f^2 h^3}-\frac{360 h'' f'}{f h^3 r}-\frac{270 h' f'}{f h^3 r^2}+\frac{540 f'}{f h r^3}-\frac{540 f'}{f h^2 r^3}+\frac{90 \left(f''\right)^2}{f^2 h^2}+\frac{180 f''}{f h^2 r^2}-\frac{180 h' f''}{f h^3 r}-\frac{180 f''}{f h r^2}+\left(-\frac{1755 \left(f'\right)^4}{8 f^4 h^2}-\frac{1035 h' \left(f'\right)^3}{4 f^3 h^3}+\frac{675 \left(f'\right)^3}{2 f^3 h^2 r}+\frac{450 h' \left(f'\right)^2}{f^2 h^3 r}+\frac{1035 f'' \left(f'\right)^2}{2 f^3 h^2}+\frac{225 h'' \left(f'\right)^2}{2 f^2 h^3}-\frac{2115 \left(h'\right)^2 \left(f'\right)^2}{8 f^2 h^4}+\frac{90 \left(f'\right)^2}{f^2 h^2 r^2}+\frac{1485 \left(h'\right)^2 f'}{2 f h^4 r}+\frac{270 h' f'}{f h^3 r^2}+\frac{495 h' f'' f'}{f^2 h^3}+\frac{585 h' h'' f'}{2 f h^4}-\frac{225 f^{(3)} f'}{f^2 h^2}-\frac{45 h^{(3)} f'}{f h^3}-\frac{315 \left(h'\right)^3 f'}{f h^5}-\frac{585 f'' f'}{f^2 h^2 r}-\frac{315 h'' f'}{f h^3 r}+\frac{270 f'}{f h r^3}-\frac{90 f'}{f h^2 r^3}+\frac{630 h'}{h^3 r^3}+\frac{855 \left(h'\right)^2 f''}{2 f h^4}+\frac{1170 h' h''}{h^4 r}+\frac{270 f^{(3)}}{f h^2 r}+\frac{90 f^{(4)}}{f h^2}-\frac{315 \left(f''\right)^2}{2 f^2 h^2}-\frac{180 f'' h''}{f h^3}-\frac{270 h' f^{(3)}}{f h^3}-\frac{180 h^{(3)}}{h^3 r}-\frac{630 h' f''}{f h^3 r}-\frac{1260 \left(h'\right)^3}{h^5 r}-\frac{180 f''}{f h^2 r^2}-\frac{270 h'}{h^2 r^3}-\frac{90}{r^4}-\frac{540}{h r^4}+\frac{630}{h^2 r^4}\right) \log f+\left(-\frac{1755 \left(f'\right)^4}{4 f^4 h^2}-\frac{1035 h' \left(f'\right)^3}{2 f^3 h^3}+\frac{675 \left(f'\right)^3}{f^3 h^2 r}+\frac{900 h' \left(f'\right)^2}{f^2 h^3 r}+\frac{1035 f'' \left(f'\right)^2}{f^3 h^2}+\frac{225 h'' \left(f'\right)^2}{f^2 h^3}-\frac{2115 \left(h'\right)^2 \left(f'\right)^2}{4 f^2 h^4}+\frac{180 \left(f'\right)^2}{f^2 h^2 r^2}+\frac{1485 \left(h'\right)^2 f'}{f h^4 r}+\frac{540 h' f'}{f h^3 r^2}+\frac{990 h' f'' f'}{f^2 h^3}+\frac{585 h' h'' f'}{f h^4}-\frac{450 f^{(3)} f'}{f^2 h^2}-\frac{90 h^{(3)} f'}{f h^3}-\frac{630 \left(h'\right)^3 f'}{f h^5}-\frac{1170 f'' f'}{f^2 h^2 r}-\frac{630 h'' f'}{f h^3 r}+\frac{540 f'}{f h r^3}-\frac{180 f'}{f h^2 r^3}+\frac{1260 h'}{h^3 r^3}+\frac{855 \left(h'\right)^2 f''}{f h^4}+\frac{2340 h' h''}{h^4 r}+\frac{540 f^{(3)}}{f h^2 r}+\frac{180 f^{(4)}}{f h^2}-\frac{315 \left(f''\right)^2}{f^2 h^2}-\frac{360 f'' h''}{f h^3}-\frac{540 h' f^{(3)}}{f h^3}-\frac{360 h^{(3)}}{h^3 r}-\frac{1260 h' f''}{f h^3 r}-\frac{2520 \left(h'\right)^3}{h^5 r}-\frac{360 f''}{f h^2 r^2}-\frac{540 h'}{h^2 r^3}-\frac{180}{r^4}-\frac{1080}{h r^4}+\frac{1260}{h^2 r^4}\right) \log\nu\right] \left(\xi -\frac{1}{6}\right)^2,
\end{dmath}
\pagebreak
\end{widetext}
and
\begin{equation}
    \langle\hat{T}^{\varphi}_{\varphi}\rangle_{\rm AHS}=\langle\hat{T}^{\theta}_{\theta}\rangle_{\rm AHS}.
\end{equation}

\section{Regularity of the Anderson-Hiscock-Samuel RSET}
\label{Appendix:AHSReg}
The expressions
(\ref{Eq:AHSRSETtt}-\ref{Eq:AHSRSETthth})
give rise to a covariantly conserved RSET obtained directly by taking the analytical byproduct of following the point-splitting renormalization procedure (as in \cite{Andersonetal1995}) or the Hadamard renormalization prescription (see \cite{BreenOttewill2012} for the application of this method to Lukewarm black hole spacetimes). The higher-derivative terms naturally arise upon isolating and subtracting all the ultraviolet divergent terms that appear in the field propagator from which the RSET is constructed. Thus, these expressions are purely geometrical and invoke no additional assumptions about the spacetime over which they are obtained, resulting in an RSET that is regular at the center of spherical symmetry, with one caveat that we detail in the following. It is straightforward to check that regularity of the Kretschmann invariant 
\begin{equation}
    \mathcal{K}=R^{\mu\nu\rho\sigma}R_{\mu\nu\rho\sigma}
\end{equation}
of the metric at $r=0$ enforces the metric functions to obey the expansions 
\begin{align}\label{Eq:MetricExpansions}
    f(r)=
    &
    a_{0}+a_{2}r^{2}+a_{3}r^{3}+a_{4}r^{4}+\order{r^{5}},\nonumber\\
    h(r)=
    &
    1+b_{2}r^{2}+b_{3}r^{3}+b_{4}r^{4}+\order{r^{5}},
\end{align}
These conditions ensure the finiteness at $r=0$ of all other invariants constructed from contractions of the Ricci and Riemann tensors and the Ricci scalar. Notice how the Kretschmann invariant constraints the value of the coefficients in \eqref{Eq:MetricExpansions} up to second-order terms in $r$ because it only involves up to second-order derivatives of the metric functions (notice the absence of linear terms in $r$ in the expansion). However, in replacing the expansions \eqref{Eq:MetricExpansions} in the AHS-RSET we obtain, for the $\langle\hat{T}^{t}_{t}\rangle^{\rm AHS}$ component
\begin{align}
1440\pi^{2}\langle\hat{T}^{t}_{t}\rangle^{\rm AHS}=
&
\frac{12a_{3}}{a_{0}r}\left[1+\log \left(a_{0}\nu^{2}\right)+60\left(\xi-\frac{1}{6}\right)\right.\nonumber\\
&
\left.
+180\log \left(a_{0}\nu^{2}\right)\left(\xi-\frac{1}{6}\right)^{2}\right]\nonumber\\
&
+\frac{4b_{3}}{r}\left[2+\log\left(a_{0}\nu^{2}\right)+60\left(\xi-\frac{1}{6}\right)\right.\nonumber\\
&
\left.-360\log\left(a_{0}\nu^{2}\right)\left(\xi-\frac{1}{6}\right)^{2}\right]
+\order{r^{0}}.
\end{align}
Notice the divergence at $r=0$ when the terms $a_{3}$ and $b_{3}$ are nonzero. Here we omit the remaining RSET components as they show similar divergences.

The presence of higher-derivative terms of the metric in the RSET imposes more restrictive conditions for regularity than those given by the finiteness of curvature invariants themselves. This adds an extra degree of non-physicality to RSET approximations that exhibit higher-derivative terms due to these terms becoming singular at $r=0$ on geometries that are entirely regular in classical general relativity. The OR-RSET is $\langle\hat{T}^{\mu}_{\nu}\rangle_{\rm OR}=\order{r^{0}}$ over any metric whose curvature invariants are finite, while also diverging at the regions where the state in which it is evaluated becomes singular, so here we advocate its use over the AHS-RSET in scenarios where the point $r=0$ belongs to the spacetime.

\bibliographystyle{unsrt}
\bibliography{biblio-semiclassical}
\end{document}